\documentclass[12pt, draftclsnofoot, onecolumn]{IEEEtran}
\usepackage[T1]{fontenc}% optional T1 font encoding
\usepackage{graphicx}
\usepackage{color}
\usepackage{titlecaps}
\usepackage{subfigure}
\usepackage{amsmath}
\usepackage{amsthm}
\usepackage{multirow}
\usepackage{amsfonts}
\usepackage{bbm}
\usepackage{dsfont}
\usepackage{epstopdf}
\usepackage{amssymb}
\newtheorem{Theorem}{Theorem}
\newtheorem{Lemma}{Lemma}
\usepackage{balance}
\usepackage{paralist}
\newcommand{\argmax}{\operatornamewithlimits{arg\,max}}
\newcommand{\argmin}{\operatornamewithlimits{arg\,min}}
\usepackage{algorithm}
\usepackage{algorithmic}
\DeclareMathOperator{\tr}{tr}

\begin{document}

\title{Content Popularity Prediction Towards Location-Aware Mobile Edge Caching}
\author{Peng~Yang,~Ning~Zhang,~Shan~Zhang,\\~Li~Yu,~Junshan~Zhang,~and~Xuemin~(Sherman)~Shen%
%\thanks{This work is supported by National Natural Science Foundation of China under Grant No. 61871437, 61801011 and the Natural Sciences and Engineering Research Council (NSERC) of Canada. Peng Yang is also financially supported by the China Scholarship Council. \emph{(Corresponding author: Li Yu.)}}%
\thanks{P. Yang and L. Yu are with the School of Electronic Information and Communications, Huazhong University of Science and Technology, Wuhan, Hubei 430074, China (e-mail: yangpeng@hust.edu.cn; hustlyu@hust.edu.cn).}%
\thanks{N. Zhang is with the Department of Computing Sciences, Texas A \& M University-Corpus Christi, Corpus Christi, TX 78412, USA (e-mail: ning.zhang@tamucc.edu).}%
\thanks{S. Zhang is with Beijing Key Laboratory of Computer Networks, School of Computer Science and Engineering, Beihang University, Beijing 100191, China (e-mail: zhangshan18@buaa.edu.cn).}%
\thanks{J. Zhang is with the School of Electrical, Computer and Energy Engineering, Arizona State University, Tempe, AZ 85287, USA (e-mail: \mbox{junshan.zhang@asu.edu}).}%
\thanks{X. Shen is with the Department of Electrical and Computer Engineering, University of Waterloo, Waterloo, ON N2L 3G1, Canada (e-mail: sshen@uwaterloo.ca).}%
\thanks{This paper was accepted in part by IEEE GLOBECOM 2017 \cite{peng1}.}}

\maketitle

\begin{abstract}
Mobile edge caching enables content delivery within the radio access network, which effectively alleviates the backhaul burden and reduces response time. To fully exploit edge storage resources, the most popular contents should be identified and cached. Observing that user demands on certain contents vary greatly at different locations, this paper devises location-customized caching schemes to maximize the total content hit rate. Specifically, a linear model is used to estimate the future content hit rate. For the case where the model noise is zero-mean, a ridge regression based online algorithm with positive perturbation is proposed. Regret analysis indicates that the proposed algorithm asymptotically approaches the optimal caching strategy in the long run. When the noise structure is unknown, an $H_{\infty}$ filter based online algorithm is further proposed by taking a prescribed threshold as input, which guarantees prediction accuracy even under the worst-case noise process. Both online algorithms require no training phases, and hence are robust to the time-varying user demands. The underlying causes of estimation errors of both algorithms are numerically analyzed. Moreover, extensive experiments on real world dataset are conducted to validate the applicability of the proposed algorithms. It is demonstrated that those algorithms can be applied to scenarios with different noise features, and are able to make adaptive caching decisions, achieving content hit rate that is comparable to that via the hindsight optimal strategy.
\end{abstract}


\section{Introduction}
The past decade has witnessed a significant growth of mobile traffic. Such growth puts tremendous pressure on the paradigm of Cloud-based service provisioning, since moving a large volume of data into and out of the cloud wirelessly requires substantial spectrum resources, and meanwhile may incur large latency. \emph{Mobile Edge Computing} (MEC) emerges as a new paradigm to alleviate the capacity concern of mobile networks \cite{ben17}. Residing on the network edge, MEC makes abundant storage and computing resources available to mobile users through low-latency wireless connections, facilitating a number of mobile services like local content caching, augmented reality, and cognitive assistance \cite{ETSI}.

Among these services, content caching at the network edge is garnering much attention \cite{fran}-\cite{icc14}. In particular, with the prevalence of social media, multimedia contents are spreading among mobile users in a viral fashion, putting high pressure on the network backhaul \cite{tsp,tit13}. It is pointed out that, by caching contents on network edge, up to $35\%$ traffic on the backhaul can be reduced \cite{ETSI}. Unfortunately, compared with the increasing content volume, the storage size at edge node (EN) is always limited. It is impossible to cache all the contents locally. Hence, identifying the optimal set of contents that maximizes cache utilization becomes crucial.

Content popularity is an effective measure for making caching decisions. Extensive works have been devoted to popularity-based content caching. According to the features of content popularity profile, those works on content caching can be classified into three categories: 1) known popularity profile \cite{tit13}-\cite{mobihoc15}; 2) fixed but unknown popularity profile \cite{tcom16, icc14}; and 3) time-varying and unknown popularity profile \cite{twc17, tmm16}. In case of fixed and unknown popularity profile, learning algorithms have been proposed under different network settings. In case of time-varying and unknown popularity profile, context information of the request, including system states and user characteristics, is exploited to make content hit rate predictions. To improve the accuracy of popularity prediction, the context space needs to be subtly designed since there is endless context information that could be taken into consideration. It is often difficult to directly identify the factors that influence content popularity. More importantly, using user information for context differentiation is subject to privacy regulations and may not be applicable in practice.

In this paper, we investigate mobile edge caching with time-varying and unknown popularity profile. Instead of relying on user information for context differentiation, we explore location features of each EN to improve the accuracy of popularity prediction, with the rationale outlined as follows. First, locations can be divided into categories with distinct social functions, such as residential area and business district. Meanwhile, users in different places have diverse interests \cite{cacm}. As indicated by real-world measurement studies \cite{jcst16}, the distribution of content popularity for even adjacent Wi-Fi APs and cellular base stations are different, and existing content caching schemes do not take such fine-grained popularity difference into consideration \cite{icme17}. To further improve the content distribution in mobile context, it is crucial to investigate content popularity with location awareness. Given that there is no established model to characterize location features and user demands, we take some initial steps to devise a model where user demand of a certain content is treated as a linear combination of content features and location characteristics with unknown noise. It follows that, the popularity prediction problem boils down to the estimation of location feature vector of each EN in the presence of noise. In practice, the noise process is affected by various factors. Firstly, it is affected by location-dependent factors, such as user interests, the number of users and the social function of the coverage area of each EN. Secondly, it is also affected by content-dependent factors, which include genre, length, and frame quality for video contents. Unfortunately, it is often difficult for content providers and edge servers to understand the statistical nature of the underlying noise process in such complicated context space. To solve the location feature estimation problem, two online prediction algorithms are proposed for different scenarios.

To start with, we consider the tractable zero-mean noise scenario as the first step. A ridge regression based prediction algorithm (RPUC) is proposed to estimate the location feature vector. To account for the impact of noise, a positive perturbation is added to the result as the correction of the prediction. By comparing to the hindsight optimal caching policy, theoretical analysis shows that the RPUC algorithm achieves sublinear regret, i.e., it asymptotically approaches the optimal strategy in the long-term.

Furthermore, we consider practical cases where noise structure is unknown \emph{a priori}. To ensure robust prediction, we resort to the $H_{\infty}$ filter technique, which enables us to obtain guaranteed accuracy even in the worst-case scenario. In particular, taking a prescribed accuracy threshold as an input, we propose an $H_{\infty}$ based prediction algorithm (HPDT), which is robust as long as the noise amplitude is finite.

Both RPUC and HPDT require no training phases, and hence are adaptive to the time-varying user demand. Numerical analysis indicates that, the regret of RPUC originates from the bias and variance of ridge regression, as well as the artificial perturbation. Note that the HPDT algorithm is conservative in that it makes no assumption on the noise. Yet, it is still able to make unbiased estimation on the location feature vector. Extensive simulations on real world traces demonstrate that those two algorithms can be applied to scenarios with different noise features, and both of them are able to make adaptive caching decisions, achieving content hit rate that is comparable to that using the hindsight optimal strategy.
The contributions of this work on mobile edge caching are three-fold:
\begin{itemize}
	\item We propose to exploit the diversity of content popularity over different locations. We establish a linear model for content popularity prediction, taking into account both content and location features. 
	\item We develop two popularity prediction algorithms that deal with different noise models. Both algorithms are able to make location-aware caching decisions. Moreover, they require no training phases, and hence can adapt to dynamic user demand.
	\item We demonstrate the effectiveness of the proposed algorithms through theoretical analysis. It is proved that performance of the RPUC algorithm asymptotically approaches that using the hindsight optimal strategy, while performance of the HPDT algorithm hinges upon noises. Experiments on real dataset crawled from YouTube show that, the long-term content hit rates of the proposed algorithms are comparable to that via the hindsight optimal strategy.
\end{itemize}

The remainder of the paper is organized as follows. Section \ref{relatedwork} reviews related works on content caching in wireless networks. Section \ref{systemmodel} describes the system model, including the mobile edge caching architecture and the formal problem formulation. In Section \ref{cachingalgorithm}, we propose the RPUC caching algorithm for the case of zero-mean noise, and give the detailed performance analysis. For the case of unknown noise model, we present the HPDT algorithm as well as detailed regret analysis in Section \ref{cachingalgorithm1}. Numerical analysis and experimental results of the two algorithms are provided in Section \ref{simulation}, followed by concluding remarks in Section \ref{conclusion}.

\section{Related Work}\label{relatedwork}
Mobile user's capacity is greatly augmented in the era of MEC. As a result, mobile service provisioning is expected to have further improved quality of experience (QoE) \cite{ETSI}. To this end, various mobile edge architectures have been proposed. Tandom \emph{et al.} proposed to deploy edge resources within radio access networks. They characterized the relationship between latency and caching size, as well as latency and fronthaul capacity, from an information-theoretic perspective \cite{fran}. Yang \emph{et al.} introduced an edge resource provisioning architecture based on cloud radio access network (C-RAN), and devised a cloud-edge interoperation scheme via software defined networking techniques \cite{peng}. Tong \emph{et al.} designed a hierarchical edge architecture, aiming at making efficient use of edge resources when serving the peak loads from mobile users \cite{infocom16}. As the 5G wireless network is expected to incorporate diverse access technologies, in this paper, we consider edge caching in the context of heterogeneous networks. Potential EN deployment can be capacity-augmented base stations, WiFi access points and other devices with excess resources.

As an effective approach to improving QoE in 5G systems, edge caching has received extensive attention \cite{commag14}. Specifically, various works have been done on video content caching, since video contents are forecast to be dominant in 5G systems \cite{nossdav16}-\cite{tmm10}. A vast amount of other works simply focus on generalized content caching. Zhang \emph{et al.} investigated the cache-enabled vehicular networks with energy harvesting, aiming at minimizing network deployment costs with QoE guarantees \cite{shan}. Ao \emph{et al.} explored distributed content caching and small cell cooperation to accelerate content delivery \cite{mobihoc15}. Device-to-device (D2D) communication is another promising solution to improve the QoE of mobile content dissemination \cite{jsac16}. Different from conventional content unicast from cellular base stations, D2D communication has the potential to significantly boost system throughput by multicasting. Ji \emph{et al.} provided a comprehensive summary on D2D caching networks, incorporating throughput scaling law and coded caching in D2D networks \cite{jsac161}. In the above works, content popularity profile was assumed to be completely known. However, in practice, content popularity may be unknown \emph{a priori}. To address this issue, various learning-based approaches have been proposed to predict content popularity. Bharath \emph{et al.} proposed a learning method that achieves desired popularity accuracy in finite training time \cite{tcom16}. Blasco \emph{et al.} modeled content caching with unknown popularity as a multi-armed bandit problem  \cite{icc14}. By carefully balancing exploration and exploitation in the learning phase, they proposed three algorithms that quickly learn content popularity under various system settings.

Unfortunately, often times content popularity profile can not only be unknown \emph{a priori}, but also time-varying. This is because user's interests change constantly, and meanwhile new contents are being created \cite{cacm}. As a result, learning-based caching algorithms should be designed in an online fashion, i.e., requiring no training phase, and adaptive to popularity fluctuations. To this end, Roy \emph{et al.} proposed to predict video popularity by utilizing knowledge from the social streams \cite{tmm13}. M\"uller \emph{et al.} introduced context-aware proactive caching \cite{twc17}. By constructing context space based on user information, they proposed an online algorithm that first learns context-specific user demands, and then updates cached contents accordingly. Other information has also been used for context differentiation, such as content features and system states \cite{tmm16}. The prediction accuracy of those solutions is highly dependent on the information used for context differentiation. To content service providers, however, user information is extremely sensitive and often unavailable. In addition, it is also impossible for them to get detailed system or network information when making caching decisions. 

In this paper, we exploit locational features for context differentiation. Locational information can be easily obtained, for example, users attached to different ENs are naturally divided into geographical groups. Based on which we investigate the location-aware caching problem with unknown and time-varying content popularity profile. By modeling user demand as linear combination of location features and content attributes, our previous work has addressed the content popularity prediction problem with the assumption that the model noise is zero-mean \cite{peng1}. As an extension, this paper additionally considers the practical scenario, where noise structure is unknown \emph{a priori}. Specifically, a robust prediction algorithm is proposed with detailed theoretical caching performance analysis. The proposed algorithm is robust and practical as it guarantees prediction accuracy regardless of the noise statistics. Additionally, numerical analysis and comparison on the root causes of estimation errors of both algorithms are presented. Much extensive experiments are conducted to validate the performance of the proposed algorithms.

It is worth noting that, in the mobile context, fetching content from the edge cache significantly reduces the delay, compared with that from conventional content distribution network (CDN). Moreover, existing content pushing strategies in CDN do not consider the fine-grained popularity differentiation in neighbouring Wi-Fi APs and cellular base stations \cite{jcst16}. With the consideration of location awareness, this paper further models and predicts the dynamics of content popularity, which is constantly varying with time.

\section{System Model and Problem Formulation}\label{systemmodel}
In this section, we present the system model and formulate the caching problem in mobile edge networks.

\subsection{Network Model}
\begin{figure}[t]
\centering
\includegraphics[width=0.55\textwidth]{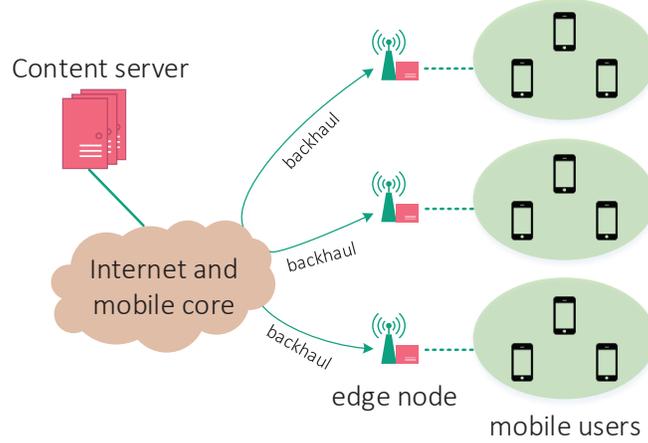}
\caption{Network model of mobile edge caching.}
\label{cachingmodel}
\end{figure}

Mobile edge computing can enhance mobile user's capacity by provisioning storage, computing and networking resources in their proximity. Capacity-augmented base stations, WiFi access points and other devices with excess capacity can be exploited for edge node deployment \cite{ben17}. In this paper, the storage resources at edge nodes are harnessed for content caching services. Specifically, as shown in Fig. \ref{cachingmodel}, a set of edge nodes $\mathcal N = \{1,2,\dotsc,N\}$ is deployed with separated backhaul links connecting to the mobile core network. Online contents are dynamically pushed to edge nodes so that user's content requests can be processed with reduced latency. Each edge node serves a disjoint set of mobile users.

\subsection{Content Popularity and Location Diversity}
\begin{figure}[t]
\centering
\includegraphics[width=0.55\textwidth]{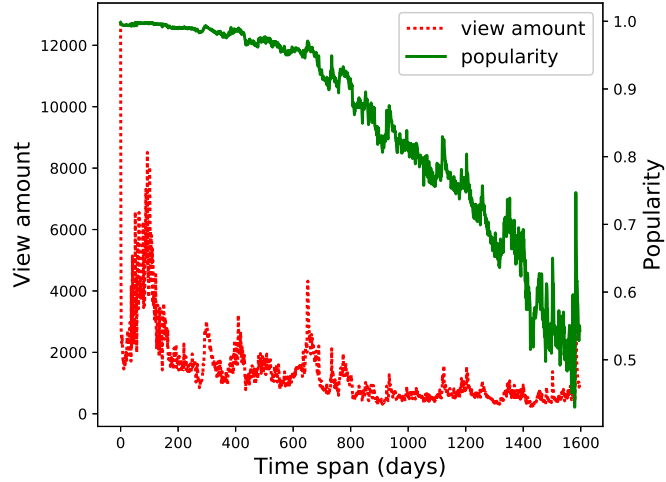}
\caption{The daily view amount and popularity trends of a YouTube video since uploaded. The popularity score equals to the ratio of the video's daily view amount to the total daily view amount of all the videos. Note the statistics are based on a set of randomly crawled videos.}
\label{Youtubestatistics}
\end{figure}

A simple yet effective caching strategy is to push the most popular contents to the network edge. Hence, local content hit rate is maximized and user's requests are served with reduced latency and improved QoE. Extensive works have been done on the popularity of contents, especially video files \cite{tsp,cacm,imc}. According to the statistics we crawled from YouTube, as illustrated in Fig. \ref{Youtubestatistics}, the popularity profile of a video file varies in two-fold. 1) The daily view amount is time-varying. 2) As other videos' daily view amounts are also varying and new videos are uploaded, the popularity of a video file is constantly fluctuating \cite{tmm15}. Moreover, location diversity also affects the content popularity. As a result, general caching strategies based on fixed popularity profile are not optimal in practice.

Let $\boldsymbol{x}_{f,n} \in \mathbb R^d$ be a $d$-dimensional attribute vector of file $f$ associated with EN $n$. For example, the attributes of video contents may include video quality, genre, length, and historical view statistics. Then, the hit rate\footnote{We define \emph{hit rate} as the number of content requests rather than a ratio.} of file $f$ at EN $n$, denoted by $d_{f,n}$, can be expressed as the following noisy linear combination
\begin{equation}\label{linearmodel}
d_{f,n} = \boldsymbol{x}_{f,n}^\top \boldsymbol{\theta}_{n}^\ast + w_{n},
\end{equation}
where $\boldsymbol{\theta}_{n}^\ast \in \mathbb R^d$ is the unknown location feature vector associated with EN $n$. Further, it also represents the location characteristics of EN $n$, which is time-invariant. $w_{n}$ is the random noise associated with EN $n$, which may be affected by various locational features, including social function of the area around EN $n$, the number of users served by EN $n$, the frequency of content update (e.g., hourly or daily). As a result, contents with the same attribute vector are expected to have different view amounts at different ENs. This linear prediction model is widely used in other areas, such as signal processing and financial engineering \cite{feng}. It provides a method to predict future hit rate and it is essential when exploiting location diversity for popular-unknown content caching. Without loss of generality, let $||\boldsymbol{\theta}_{n}^\ast|| \leq \zeta$, $||\boldsymbol{x}_{f,n}|| \leq \eta$ and  $d_{f,n} \leq \gamma$ for all $f \in \mathcal F$ and $n \in \mathcal N$,  where $||\boldsymbol{x}|| = \sqrt{\boldsymbol{x}^\top\boldsymbol{x}}$ denotes the Euclidean norm of $\boldsymbol{x}$, $\zeta$, $\eta$ and $\gamma$ are positive constants. Also, for notational simplicity, define $||\boldsymbol{x}||_{\boldsymbol{V}} \triangleq \sqrt{\boldsymbol{x}^\top \boldsymbol{V}\boldsymbol{x}}$ as the weighted (by a matrix $\boldsymbol{V} \in \mathbb R^{d \times d}$) Euclidean norm of $\boldsymbol{x}$.

\subsection{Problem Formulation}
Consider a set of files $\mathcal F = \{1,2,\dotsc,F\}$ that can be cached at ENs, and let $c<F$ be the caching size of each EN. We assume that all the contents are of equal size\footnote{In case contents are of different sizes, they are split into smaller ones of equal size. For example, the widely used DASH (Dynamic Adaptive Streaming over HTTP) protocol breaks contents into small segments before transmission. This assumption is used to simplify the theoretical analysis, and a similar assumption has been made in \cite{tmm16, twc16}. Location-aware edge caching with different content sizes deserves further investigation.} and the size is normalized to 1, i.e., each EN can cache up to $c$ contents.
As indicated by Fig. \ref{Youtubestatistics}, content popularity is time-varying. Therefore, contents with higher popularity should be proactively identified and cached at the ENs, and the less popular ones should be evicted so as to improve the local content hit rate. Considering a sequence of time slots $\mathcal T = \{1,2,\dotsc,T\}$, and let $\mathcal F_{n,t}$ denote the set of contents cached at EN $n$ during time slot $t \in \mathcal T$, and $d_{f,n,t}$ be the amount of user demand on file $f$ at EN $n$ during time slot $t$. The objective of a caching policy is to maximize the time-averaged hit rate. Formally, it can be formulated as the following time-averaged hit rate maximization (THRM) problem\footnote{Without loss of generality, we assume that the underlying process is ergodic.}:
\begin{equation}
\begin{array}{rl}
\mbox{(\underline{THRM}):}\max & \! \frac{1}{T} \sum_{t\in\mathcal T} \sum_{n \in \mathcal N} \sum_{f \in \mathcal F_{n,t}} d_{f,n,t} \\
\textrm{Subject to:} & \!|\mathcal F_{n,t}| \leq c, \; \forall n\in\mathcal N, \; t \in \mathcal T.\label{LHRM}
\end{array}
\end{equation}
As the amount of user demand, i.e., the hit rate of contents at each EN, is unknown \emph{a priori}, the decision variables $\mathcal F_{n,t}$ in problem (\ref{LHRM}) is intractable directly. For convenience, denoting the optimal caching strategy $\mathcal F_{n,t}^\ast$ for EN $n$ at time $t$, we have
\begin{equation}\label{optimalset}
\mathcal F_{n,t}^\ast = \argmax_{|\mathcal F_{n,t}| \leq c}\sum_{f \in \mathcal F_{n,t}}d_{f,n,t},\; \forall n \in \mathcal N,\; t \in \mathcal T.
\end{equation}
Define the time-averaged caching \emph{regret} of a solution respect to the optimal caching strategy as
\begin{equation}\label{regret}
R(T) \triangleq \frac{1}{T} \, \mathbb E\left[ \sum_{t\in\mathcal T} \sum_{n \in \mathcal N} \left(\sum_{f \in \mathcal F_{n,t}^\ast} d_{f,n,t} - \sum_{f \in \mathcal F_{n,t}} d_{f,n,t}\right) \right].
\end{equation}
Then, the THRM problem can be reformulate as a time-averaged \emph{regret} minimization (TRM) problem:
\begin{equation}
\begin{array}{rl}
\mbox{(\underline{TRM}):}\min & R(T)\\
\textrm{Subject to:} & \!|\mathcal F_{n,t}| \leq c, \; \forall n\in\mathcal N, \; t \in \mathcal T.\label{LRM}
\end{array}
\end{equation}
Given that the optimal set $\mathcal F_{n,t}^\ast$ is unknown \emph{a priori}, our goal is to develop a caching policy that constantly makes good estimation of the optimal set $\mathcal F_{n,t}^\ast$, and therefore minimizes the time-averaged caching regret. As indicated by Eq. (\ref{optimalset}), the LRM problem boils down to estimating user demands of different contents at each EN. Given the linear model in Eq. (\ref{linearmodel}), if the location feature vector $\boldsymbol{\theta}_{n}^{\ast}$ can be found in the presence of noise, we can make an accurate prediction on user demand. Unfortunately, there is no established statistical model on the noise processes that impinges the prediction of user demand. In what follows, we propose two online content popularity prediction algorithms by making dynamic estimations on the location feature vectors for different noise processes. In particular, the first algorithm achieves near-optimal performance with the assumption that the model noise is zero-mean, while the second algorithm is designed to provide robust performance guarantees in the case of unknown noise statistics.

\section{Ridge Regression based Content Popularity Prediction and Edge Caching}\label{cachingalgorithm}
In this section, as the first attack on the TRM problem, we present a caching algorithm when noise is zero-mean. 

\subsection{Location Feature Vector Estimation}
When the noise is zero-mean, according to Eq. (\ref{linearmodel}), we have
\begin{equation}\label{linearprediction}
\mathbb E[d_{f,n}|\boldsymbol{x}_{f,n}] = \boldsymbol{x}_{f,n}^\top \boldsymbol{\theta}_{n}^\ast.
\end{equation}
It can be interpreted that, at time slot $t$, given the attribute vector $\boldsymbol{x}_{f,n,t}$, the hit rate of file $f$ at EN $n$ is predicted to be the linear combination of its attributes, which provides a feasible way to predict the content hit rate. Since the location feature vector $\boldsymbol{\theta}_n^\ast$  of EN $n$ is time-invariant, a good estimation of $\boldsymbol{\theta}_n^\ast$ will lead to accurate prediction of the content hit rate.

Let the attribute matrix $\boldsymbol{\Phi}_{f,n}\in \mathbb R^{m \times d}$ be the historical data up to time slot $t$, where $m$ is the frequency of file $f$ being cached at EN $n$ up to time slot $t$, and the $m$-th row of $\boldsymbol{\Phi}_{f,n}$ is the corresponding attribute vector $\boldsymbol{x}_{f,n,m}$. Denote by $\boldsymbol{y}_{f,n}\in\mathbb R^m$ the $m$-time empirical hit rate of file $f$ at EN $n$. By applying the standard ordinary least square linear regression, i.e., $\boldsymbol{\theta}_n^\ast = {\argmin}_{\boldsymbol{\theta}_n} ||\boldsymbol{y}_{f,n} - \boldsymbol{\Phi}_{f,n}\boldsymbol{\theta}_n||^{2}$, we can obtain the unique solution $\boldsymbol{\theta}_n^\ast = (\boldsymbol{\Phi}_{f,n}^\top \boldsymbol{\Phi}_{f,n})^{-1}\boldsymbol{\Phi}_{f,n}^\top \boldsymbol{y}_{f,n}$, which is unbiased. However, when there are correlated variables in the attribute vector, the matrix $\boldsymbol{\Phi}_{f,n}^\top \boldsymbol{\Phi}_{f,n}$ may not be invertible. As a result, the estimated $\boldsymbol{\theta}_n$ can be poorly determined and will exhibit high variance.

In contrast to the unbiased estimation, ridge regression makes biased estimation by adding a control parameter that ``penalizes'' the magnitude of estimated $\boldsymbol{\theta}_n$, which helps to improve estimation stability. Specifically, ridge regression aims at minimizing a penalized sum
\begin{equation}
\boldsymbol{\theta}_n^\ast = \argmin_{\boldsymbol{\theta}_n} \Big(||\boldsymbol{y}_{f,n} - \boldsymbol{\Phi}_{f,n}\boldsymbol{\theta}_n||^{2} + \mu ||\boldsymbol{\theta}_n||^{2} \Big)\,,
\end{equation}
where $\mu > 0$ controls the size of $\boldsymbol{\theta}_n$: the larger the value of $\mu$, the greater the shrinkage of the magnitude of $\boldsymbol{\theta}_n$ \cite{bookstatistic}. Consequently, the estimation of $\boldsymbol{\theta}_n^\ast$ can be explicitly given as
\begin{equation}\label{ridge}
\tilde{\boldsymbol{\theta}}_n = (\boldsymbol{\Phi}_{f,n}^\top \boldsymbol{\Phi}_{f,n} + \mu \boldsymbol{I}_d)^{-1} \boldsymbol{\Phi}_{f,n}^\top \boldsymbol{y}_{f,n}\,,
\end{equation}
where $\boldsymbol{I}_d \in \mathbb R^{d \times d}$ is the identity matrix. The accuracy of the estimation depends on the amount of data and the selection of $\mu$. For convenience, let $\boldsymbol{V}_{f,n} = \boldsymbol{\Phi}_{f,n}^\top \boldsymbol{\Phi}_{f,n} + \mu \boldsymbol{I}_d$ for all $f \in \mathcal F$ and $n \in \mathcal N$. The following lemma, which is slightly manipulated from \cite{chu}, gives an upper bound on the estimation error of ridge regression.

\begin{Lemma}\label{estimationerror}
If $||\boldsymbol{\theta}_{n}^\ast|| \leq \zeta$ for all $n \in \mathcal N$, then $\forall  \delta >0$, the estimation error of ridge regression can be upper bounded as 
\begin{equation}\label{errorbound}
|\boldsymbol{x}_{f,n}^\top \tilde{\boldsymbol{\theta}}_n - \boldsymbol{x}_{f,n}^\top \boldsymbol{\theta}_n^\ast| \leq (\delta + \zeta\mu) ||\boldsymbol{x}_{f,n}||_{\boldsymbol{V}_{f,n}^{-1}}
\end{equation}
with probability at least $1-2e^{-2\delta^2}$.
\end{Lemma}

Please refer to Appendix \ref{prooflemma1} for the proof. The probabilistic upper bound of estimation error provided in Lemma \ref{estimationerror} indicates that, the true hit rate $\boldsymbol{x}_{f,n}^\top \boldsymbol{\theta}_n^\ast$ falls into the confidence interval around the estimation $\boldsymbol{x}_{f,n}^\top \tilde{\boldsymbol{\theta}}_n$ with high probability. The righthand side of Eq. (\ref{errorbound}) gives the length of the confidence interval, which is crucial to the following content popularity prediction and caching algorithm. 

\subsection{RPUC Caching Algorithm}

\begin{algorithm}[t]
\caption{RPUC: \textbf{R}idge Regression \textbf{P}rediction with \textbf{U}pper \textbf{C}onfidence for Location-Aware Edge Caching}
\begin{algorithmic}[1] \label{RPUC}
\REQUIRE $\mu >0$.
\ENSURE Set of files to be cached in each EN.
\STATE Initialization: Cache files in every EN and get the initial attribute vectors $\boldsymbol{x}_{f,n,0}$ of all file-EN pairs.
\STATE $\boldsymbol{V}_{n} \gets \mu \boldsymbol{I}_d$, $\boldsymbol{h}_{n} \gets \boldsymbol{0}_d$, $\forall n \in \mathcal N$
\FOR {$t = 1,2,\dotsc,T$}
\FOR {each EN $n \in \mathcal N$}
\STATE $\tilde{\boldsymbol\theta}_{n,t} \gets \boldsymbol{V}_{n}^{-1} \boldsymbol{h}_{n}$
\FOR {each file $f \in \mathcal F$}
\STATE Obtain attribute vector $\boldsymbol{x}_{f,n,t}$
\STATE $\tilde d_{f,n,t} \gets \boldsymbol{x}_{f,n,t}^\top\tilde{\boldsymbol\theta}_{n,t}$
\STATE Compute the perturbation $p_{f,n,t}$ in Eq. (\ref{perturbation})
\STATE  $\hat d_{f,n,t} \gets \tilde d_{f,n,t} + p_{f,n,t}$
\ENDFOR
\STATE $\mathcal F_{n,t} = \argmax_{\mathcal F_n \subseteq \mathcal F,\;|\mathcal F_n| \leq c} \sum_{f\in\mathcal F_n} \hat d_{f,n,t}$
\STATE Cache all the files in set $\mathcal F_{n,t}$ on EN $n$
\STATE Observe empirical demands $d_{f,n,t}$ of cached files
\STATE Update $\boldsymbol{V}_{n}$ and $\boldsymbol{h}_{n}$ based on $\boldsymbol{x}_{f,n,t}$ and $d_{f,n,t}$ of all cached files: \\
		$\boldsymbol{V}_{n} \gets \boldsymbol{V}_{n} + \boldsymbol{x}_{f,n,t}\boldsymbol{x} _{f,n,t}^\top$ \\
		$\boldsymbol{h}_{n} \gets \boldsymbol{h}_{n} + \boldsymbol{x}_{f,n,t}d_{f,n,t}$
\ENDFOR
\ENDFOR
\end{algorithmic}
\end{algorithm}

The location-aware edge caching algorithm is sketched in Algorithm \ref{RPUC}. After initialization, the algorithm iteratively performs the following three phases.

\begin{enumerate}[1.]
\item \emph{Predict}: During each time slot $t$, the location feature vector $\tilde{\boldsymbol\theta}_{n,t}$ is firstly updated according to the demand information observed in time slot $t-1$. Then, based on the linear prediction model, the estimated demand $\tilde d_{f,n,t}$ is obtained. Considering the impact of random noises, a perturbation is added to the estimation, i.e., the ultimate hit rate is predicted to be 
\begin{equation}
\hat d_{f,n,t} = \boldsymbol{x}_{f,n,t}^\top\tilde{\boldsymbol\theta}_{n,t} + p_{f,n,t} \,,
\end{equation}
where the perturbation $p_{f,n,t}$ is given by
\begin{equation}\label{perturbation}
p_{f,n,t} = \alpha_t ||\boldsymbol{x}_{f,n,t}||_{\boldsymbol{V}_{f,n}^{-1}},
\end{equation}
and $\alpha_t = \big[\ln (tF^{\frac{1}{2}})\big]^{\frac{1}{2}} + \mu\zeta$.

\item \emph{Optimize and cache}: Based on the predicted hit rate $\hat d_{f,n,t}$ of each content, a set of contents $\mathcal F_{n,t}$ that maximizes the content hit rate at EN $n$ during time slot $t$ is identified and cached respectively. Note that, certain contents may be cached in multiple ENs simultaneously.

\item \emph{Observe and update}: At the end of time slot $t$, the empirical hit rate information of cached files $\mathcal F_{n,t}$ on each EN is recorded, which is then used to update the parameter matrices for subsequent estimation and prediction.
\end{enumerate}

The rationale of the perturbation is that Eq. (\ref{linearprediction}) only gives a mean value of the hit rate which omits the potential random fluctuation, while Lemma \ref{estimationerror} provides a probabilistic upper bound  of the demand estimation error. The perturbation given in Eq. (\ref{perturbation}) is inline with the righthand side of Eq. (\ref{errorbound}) and can be regarded as the optimism in face of uncertainty, or equivalently, the upper confidence of the demand estimation. By adding a perturbation according to Eq. (\ref{perturbation}), we have $\delta =  \big[\ln (tF^{\frac{1}{2}})\big]^{\frac{1}{2}}$.  According to Lemma \ref{estimationerror}, the upper bound holds with probability at least $1-2F^{-1}t^{-2}$, which approximates to $1$ rapidly as $t$ increases.

\subsection{Regret Analysis}\label{regretanalysis}
The content hit rate of the RPUC algorithm highly depends on the accuracy of prediction. This subsection gives a theoretical upper bound on its time-averaged caching regret $R(T)$.

In mobile edge caching, let $c$ be the caching size of each EN, and $F$ be the size of ground file set. Note that content hit rate satisfies the linear model, and content attribute vectors and user demands are bounded by  $||\boldsymbol{x}_{f,n,t}|| \leq \eta$ and $d_{f,n,t} \leq \gamma$ for all $f \in \mathcal F$, $n \in \mathcal N$ and $t\in\mathcal T$, we have the following theorem.

\begin{Theorem}\label{theorem1}
If the noise is zero-mean, the RPUC algorithm achieves near-optimal performance, i.e., the time-averaged regret $R(T)$ is of order $O\Big(cN\sqrt{\frac{d (\ln T) \ln(\mu + T\eta^2/d)}{T}}\Big)$, and $R(T)\to 0$ when $T\to \infty$.
\end{Theorem}
The proof has been relegated to Appendix \ref{prooftheorem1}. Basically, the root-cause of regret is two-fold: the estimation error and the perturbation. Particularly, the estimation error consists of the linear model error and the intended bias incurred by $\mu>0$ in ridge regression. The perturbation term is well managed by the time-varying control parameter $\alpha_t$. Theorem \ref{theorem1} indicates that under RPUC algorithm, the content hit rate asymptotically approaches the optimal caching policy in the long term.

\section{Robust Content Popularity Prediction and Edge Caching}\label{cachingalgorithm1}
In the previous section, we proposed the online caching algorithm RPUC based on the linear prediction model given in Eq. (\ref{linearprediction}). A biased estimation of the location feature vector $\boldsymbol{\theta}_{n}^\ast$ for each EN is obtained by ridge regression. Further, a perturbation is added to the estimation of content hit rate to account for uncertainty. However, this caching algorithm would not work well when the noise is not zero-mean. Even worse, it is likely that we are unable to get detailed noise statistics, as it is affected by various location features, such as population, social function or even weather condition. Therefore, robust prediction algorithm that could handle noise uncertainty is desirable. In this section, by resorting to the $H_{\infty}$ filter technique, we propose a popularity prediction algorithm that provides guaranteed accuracy in the case of unknown noise structures.

\subsection{Noisy Model for Content Popularity}
Denoted by $w_{n,t}$ the additive noise added to the linear model. The linear model is rewritten as
\begin{equation}\label{newlinearmodel}
d_{f,n,t} = \boldsymbol{x}_{f,n,t}^\top \boldsymbol{\theta}_{n}^\ast + w_{n,t}.
\end{equation}
If the noise process $w_{n,t}$ follows white Gaussian distribution and its mean and correlation are always known, Kalman filtering technique can be applied to estimate $\boldsymbol{\theta}_{n}^\ast$, which achieves the smallest possible standard deviation of the estimation error \cite{shen}-\cite{simon}. Since there is no established model on the statistics of the noise structure, robust estimators on the location feature vector that can tolerate noise uncertainty are needed. Next, we will introduce the $H_\infty$ filtering technique for location feature vector estimation, which requires no \textit{a priori} information on the noise process. The only assumption is that, the magnitude of the noise process is finite, which is true since the total demand $d_{f,n,t}$ is always finite in reality.

\subsection{An $H_\infty$ Filter Approach}
When locational features are time-invariant, the true location feature vector $\boldsymbol{\theta}_{n}^\ast$ remains the same across the time span. For notational simplicity, in this subsection, we focus on a specific content on a certain EN and neglect indices $f$ and $n$. Based on Eq. (\ref{newlinearmodel}), the location feature vector estimation problem is reformulated as
\begin{equation}\label{measure}
\Bigg \{\!\!
\begin{array}{rcl}
\boldsymbol{\theta}_{t+1} & = & \boldsymbol{\theta}_{t} \\
d_{t} &=& \boldsymbol{x}_{t}^\top \boldsymbol{\theta}_{t} + w_{t}.
\end{array}
\end{equation}
Different from Kalman filter, we aim at providing a uniformly small estimation error, $e_{t} = |d_{t} - \tilde d_{t}|$, for any form of noise process. Notice that estimation of the location feature vector $\boldsymbol{\theta}^\ast$ is crucial to minimizing the estimation error $e_{t}$. Then, we define the following cost function of estimation \cite{shen}:
\begin{equation} \label{costfunction}
\mathcal J_0 \triangleq \frac{\sum_{t=0}^{T}||\boldsymbol{\theta}^\ast-\tilde{\boldsymbol{\theta}}_{t}||^2}{||\boldsymbol{\theta}_{0}-\tilde{\boldsymbol{\theta}}_{0}||_{\boldsymbol{P}_0}^2 + \sum_{t=0}^{T}|w_{t}|^2},
\end{equation}
where $\boldsymbol{P}_0$ is a symmetric positive definite matrix reflecting the confidence of the \textit{a priori} knowledge of the initial state. A 'smaller' choice of $\boldsymbol{P}_0$ indicates larger uncertainty of the initial condition and vice versa. The objective is to make a sequence of estimation on $\tilde{\boldsymbol{\theta}}_{t}$ such that the above cost is minimized. The denominator of the cost function can be regarded as a combined norm of all possible initial states and noises affecting the system. Given that there is no established stochastic model for $w_{t}$, the cost function in Eq. (\ref{costfunction}) allows us to make robust estimation of content popularity from game-theoretical perspective. Suppose that there is an unrestricted adversary, who can control the initial state $\boldsymbol{\theta}_{0}$ and the magnitude of $w_{t}$ to maximize the error of our estimation. While we focus on minimizing the numerator of Eq. (\ref{costfunction}), the adversary may incur infinite magnitude of disturbances. The form of $\mathcal J_0$ prevents the adversary from using brute force to maximize $||\boldsymbol{\theta}^\ast-\tilde{\boldsymbol{\theta}}_{t}||$. Instead, the adversary needs to carefully choose $\boldsymbol{\theta}_{0}$ and $w_{t}$ as it tries to maximize $||\boldsymbol{\theta}^\ast-\tilde{\boldsymbol{\theta}}_{t}||$. Formally, this game can be generalized as a minmax problem, and the optimal estimation on $\tilde{\boldsymbol{\theta}}_{t}$ achieves the minimal cost $ \mathcal J_0^\ast$ as
\begin{equation}
\mathcal J_0^\ast = \min_{\tilde{\boldsymbol{\theta}}_{t}}\, \max_{\boldsymbol{\theta}_{0},\,w_{t}}\, \mathcal J_0.
\end{equation}
Given the cost function in Eq. (\ref{costfunction}), directly minimizing $\mathcal J_0$ is challenging. In practice, a better approach for $H_\infty$ filtering is to seek a sub-optimal estimation that meets a given threshold. Specifically, one can try to find $\tilde{\boldsymbol{\theta}}_{t}$ such that the optimal estimate of $\boldsymbol{\theta}_{t}$ among all possible $\tilde{\boldsymbol{\theta}}_{t}$ (even including the worst-case performance measure) should satisfy
\begin{equation}\label{costthreshold}
\sup \, \mathcal J_0 < \psi^2,
\end{equation}
where $\psi>0$ is the prescribed performance bound. Eq. (\ref{costthreshold}) indicates that $H_\infty$ filter guarantees the smallest estimation error over all possible finite disturbances of the noise magnitude \cite{shen}. Rearranging Eq. (\ref{costthreshold}) gives the following equivalent minmax problem
\begin{eqnarray}
\min_{\tilde{\boldsymbol{\theta}}_{t}}\, \max_{\boldsymbol{\theta}_{0},\,w_{t}}\,\,\mathcal J &\!\!\!\triangleq\!\!\!& -\,\,\frac{1}{2}\psi^2||\boldsymbol{\theta}_{0}-\tilde{\boldsymbol{\theta}}_{0}||_{\boldsymbol{P}_0}^2 \nonumber \\
&&+\,\, \frac{1}{2}\sum_{t=0}^T \Big[||\boldsymbol{\theta}^\ast-\tilde{\boldsymbol{\theta}}_{t}||^2 - \psi^2|w_{t}|^2\Big].\label{newcostfunction}
\end{eqnarray}
This minmax problem can be interpreted as a zero-sum game against the adversary. With a given $\psi$, our goal is to find an estimation that wins the game (i.e., achieve a negative cost $\mathcal J<0$).  By resorting to the Lagrange multiplier method for the dynamic constrained optimization problem (\ref{newcostfunction}), the $H_\infty$ filter approach results in the following iterative algorithm to find the optimal estimates $\tilde{\boldsymbol{\theta}}_{t}$ for all $t\in\mathcal T$:
\begin{equation}
\tilde{\boldsymbol{\theta}}_{t+1} = \tilde{\boldsymbol{\theta}}_{t} + \boldsymbol R_t(d_{t}-\boldsymbol{x}_{t}^\top \tilde{\boldsymbol{\theta}}_{t}),
\end{equation}
where $\tilde{\boldsymbol{\theta}}_{t}$ is initialized as $\tilde{\boldsymbol{\theta}}_{0} = \boldsymbol{0}_{d}$, $\boldsymbol R_t$ is the $H_\infty$ filter gain, given by
\begin{equation}
\boldsymbol R_t = \boldsymbol M_t \boldsymbol{x}_{t}(1+||\boldsymbol{x}_{t}||_{\boldsymbol M_t}^2)^{-1},
\end{equation}
and
\begin{equation}\label{MInverse}
\boldsymbol M_{t+1}^{-1} = \boldsymbol M_{t}^{-1} + \boldsymbol{x}_{t}\boldsymbol{x}_{t}^\top - \psi^{-2}\boldsymbol{I}_d,
\end{equation}
with $\boldsymbol M_{t}^{-1}$ initialized by $\boldsymbol M_{0}^{-1} = {\boldsymbol{P}_0}- \psi^{-2}\boldsymbol{I}_d$. The detailed proof of this solution can be found in \cite{simon}. With the aid of $H_\infty$ filter, we are able to make performance-guaranteed estimation on the location feature vector $\boldsymbol{\theta}^\ast$, regardless of the detailed noise structure. Based on the estimation of location feature vector, content popularity prediction and caching algorithm can be further devised.

\begin{algorithm}[t]
\caption{HPDT : $\boldsymbol{H}_\infty$ filter \textbf{P}rediction with \textbf{D}ynamic \textbf{T}hreshold for Location-aware Edge Caching}
\begin{algorithmic}[1] \label{HPDT}
\REQUIRE $\xi$ close to but larger than $1$.
\ENSURE Set of files to be cached in each EN.
\STATE Initialization: Cache files in every EN and get the initial attribute vectors $\boldsymbol{x}_{f,n,0}$ of all file-EN pairs;\\
	Choose symmetric positive definite $\boldsymbol{P}_{n,0} \succ 0$ for all $n\in\mathcal N$;\\
	Choose the smallest possible value of $\psi_{n,0}$, so that ${\boldsymbol{P}_{n,0}}- \psi_{n,0}^{-2}\boldsymbol{I}_d$ is nonsingular;\\
	$\boldsymbol M_{n,0}^{-1} \gets {\boldsymbol{P}_{n,0}}- \psi_{n,0}^{-2}\boldsymbol{I}_d$,
	$\tilde{\boldsymbol\theta}_{n,0} \gets \boldsymbol{0}_d$ for all $n\in\mathcal N$.
\FOR {$t = 1,2,\dotsc,T$}
\FOR {each EN $n \in \mathcal N$}
\FOR {each file $f \in \mathcal F$}
\STATE Obtain the attribute vector $\boldsymbol{x}_{f,n,t}$
\STATE Compute the estimated user demand:\\ $\tilde d_{f,n,t} \gets \boldsymbol{x}_{f,n,t}^\top\tilde{\boldsymbol\theta}_{n,t}$
\ENDFOR
\STATE $\mathcal F_{n,t} \gets \argmax_{\mathcal F_n \subseteq \mathcal F,\;|\mathcal F_n| \leq c} 			\sum_{f\in\mathcal F_n} \tilde d_{f,n,t}$
\STATE Cache the set of files $\mathcal F_{n,t}$ in EN $n$
\STATE Observe user demand $d_{f,n,t}$ of cached files
\STATE Update $\tilde{\boldsymbol{\theta}}_{n,t}$ based on $\boldsymbol{x}_{f,n,t}$ and $d_{f,n,t}$ of all cached files: $\tilde{\boldsymbol\theta}_{n,t} \gets \tilde{\boldsymbol\theta}_{n,t-1} + \boldsymbol R_{n,t}(d_{f,n,t}-\boldsymbol{x}_{f,n,t-1}^\top \tilde{\boldsymbol{\theta}}_{n,t-1})$, where\\
	$\boldsymbol R_{n,t} = \boldsymbol M_{n,t} \boldsymbol{x}_{f,n,t}(1+||\boldsymbol{x}_{f,n,t}||_{\boldsymbol M_{n,t}}^2)^{-1}$\\
	$\boldsymbol M_{n,t+1}^{-1} = \boldsymbol M_{n,t}^{-1} + \boldsymbol{x}_{f,n,t}\boldsymbol{x}_{f,n,t}^\top - \psi_{n,t+1}^{-2}\boldsymbol{I}_d$\\
	$\psi_{n,t+1}^{-2} = \xi^{-2}\lambda_{\min}(\boldsymbol M_{n,t}^{-1} + \boldsymbol{x}_{f,n,t}\boldsymbol{x}_{f,n,t}^\top)$
\ENDFOR
\ENDFOR
\end{algorithmic}
\end{algorithm}

\subsection{From Determining $\psi$ to the HPDT Caching Algorithm}
The performance of $H_\infty$ filter highly depends on the prescribed threshold $\psi$. A smaller $\psi$ results in a smaller estimation error. However, if $\psi$ is too small, $\boldsymbol M_{t}^{-1} + \boldsymbol{x}_{t}\boldsymbol{x}_{t}^\top - \psi^{-2}\boldsymbol{I}_d$ in  Eq. (\ref{MInverse}) may be singular, which renders the iterative solution infeasible. Hence, the value of $\psi$ should be carefully selected.
An adaptive scheme for threshold selection was proposed in \cite{shen}, which makes online iterative prediction possible. Denote by $\psi_{t+1}$ the threshold on the $(t+1)$-th iteration, it should be properly chosen to guarantee $\boldsymbol M_{t+1}^{-1}$ is positive definite, i.e.,
\begin{equation}
\boldsymbol M_{t}^{-1} + \boldsymbol{x}_{t}\boldsymbol{x}_{t}^\top - \psi_{t+1}^{-2}\boldsymbol{I}_d \succ 0.
\end{equation}
Denote by $\lambda_{\max}(\boldsymbol{A}) = \lambda_1(\boldsymbol{A}) \geq \cdots \geq \lambda_k(\boldsymbol{A}) \cdots \geq \lambda_d(\boldsymbol{A}) = \lambda_{\min}(\boldsymbol{A})$ the eigenvalues of $d \times d$ matrix $\boldsymbol{A}$, and $\lambda_k(\boldsymbol{A})$ is the $k$-th largest eigenvalue. According to the min-max theorem on matrix eigenvalues, the adaptive threshold $\psi_{t+1}$ should satisfy
\begin{equation}
\lambda_k(\boldsymbol M_{t}^{-1} + \boldsymbol{x}_{t}\boldsymbol{x}_{t}^\top) > \lambda_k(\psi_{t+1}^{-2}\boldsymbol{I}_d) \,,
\end{equation}
for all $k \in \{1,2,\cdots,d\}$. Since $\lambda_k(\psi_{t+1}^{-2}\boldsymbol{I}_d) = \psi_{t+1}^{-2}$ holds for all $k$, equivalently, we have
$\lambda_{\min}(\boldsymbol M_{t}^{-1} + \boldsymbol{x}_{t}\boldsymbol{x}_{t}^\top) > \psi_{t+1}^{-2}$.
We may let
\begin{equation}\label{psivalue}
\psi_{t+1}^{-2} = \xi^{-2}\lambda_{\min}(\boldsymbol M_{t}^{-1} + \boldsymbol{x}_{t}\boldsymbol{x}_{t}^\top),
\end{equation}
where $\xi$ is a constant very close to but larger than one, so that $\lambda_{\min}(\boldsymbol M_{t}^{-1} + \boldsymbol{x}_{t}\boldsymbol{x}_{t}^\top - \psi_{t+1}^{-2}\boldsymbol{I}_d)$ is guaranteed to be positive and hence $\boldsymbol M_{t+1}^{-1}$ is nonsingular. Meanwhile, the magnitude of $\psi_{t}$ is also suppressed.

With the aid of $H_\infty$ filter technique, we are able to make performance-guaranteed estimation on the location feature vectors. Therefore, more precise prediction on content popularity can be made, and hence differentiated caching policies can be devised on each EN. The corresponding prediction and caching algorithm is sketched in Algorithm \ref{HPDT}. Similar to Algorithm \ref{RPUC}, the iteration of the HPDT algorithm can be generalized into the following three steps after initialization.

\begin{enumerate}[1.]
\item \emph{Predict}: During time slot $t$, estimation of the location feature vector $\boldsymbol{\theta}_{n}^\ast$ of each EN is predicted based on the updated location feature vector by the end of time slot $t-1$.
\item \emph{Optimize and cache}: Based on the predicted content hit rate profile, the set of contents with maximized predicted content hit rate are cached on each EN respectively. Certain contents may be cached in multiple ENs simultaneously.
\item \emph{Observe and update}: At the end of time slot $t$, the empirical hit rate of the cached files on each EN is observed, which is then used to update the input of the $H_{\infty}$ filtering process, yielding the updated location feature vector.
\end{enumerate}

The adaptive adjustment of $\psi$ in Eq. (\ref{psivalue}) is crucial to the online HPDT algorithm. It is tuned to its minimum at each iteration, so that $\boldsymbol M_{t}$ is guaranteed to be positive definite, and meanwhile the upper bound of the cost function $J_0$ is minimized.

\subsection{Regret Analysis}
The $H_\infty$ filter technique provides a robust estimation of the location feature vector regardless of the statistical model of the noise process. However, this approach is also conservative since it needs to accommodate the disturbances of all kinds of noise processes. In this subsection, the performance bound of $H_\infty$ filter based prediction and caching algorithm is given.

Note that the prescribed performance threshold is crucial to the prediction accuracy, the adaptive threshold $\psi_{t}$ in HPDT algorithm is firstly characterized in this subsection. Note that $\boldsymbol{P}_0$ is initialized as a $d \times d$ symmetric and positive definite matrix, and $\boldsymbol M_{t}^{-1}$ is also guaranteed to be symmetric and positive definite with the help of $\psi_{t}$. According to Weyl's monotonicity theorem \cite{simax}, the smallest eigenvalue of $\boldsymbol M_{t}^{-1} + \boldsymbol{x}_{t}\boldsymbol{x}_{t}^\top$ can be bounded as:
\begin{equation}
\lambda_d(\boldsymbol M_{t}^{-1}) \leq \lambda_d(\boldsymbol M_{t}^{-1} + \boldsymbol{x}_{t}\boldsymbol{x}_{t}^\top) \leq \lambda_d(\boldsymbol M_{t}^{-1}) + \lambda_1(\boldsymbol{x}_{t}\boldsymbol{x}_{t}^\top),
\end{equation}
where $\lambda_1(\boldsymbol{x}_{t}\boldsymbol{x}_{t}^\top)$ is the largest eigenvalue of $\boldsymbol{x}_{t}\boldsymbol{x}_{t}^\top$, and can be simple bounded by matrix trace as $\lambda_1(\boldsymbol{x}_{t}\boldsymbol{x}_{t}^\top) \leq \tr(\boldsymbol{x}_{t}\boldsymbol{x}_{t}^\top) = ||\boldsymbol{x}_{t}||^2 \leq \eta^2$. As $\boldsymbol M_{t+1}^{-1}$ is positive definite, we have
\begin{equation}
\xi^{-2}\lambda_d(\boldsymbol M_{t}^{-1}) \leq \psi_{t+1}^{-2} \leq \xi^{-2}(\lambda_d(\boldsymbol M_{t}^{-1})+\eta^{2})\,,
\end{equation}
where $\xi^{-2}$ is very close to but smaller than one. According to Eq. (\ref{MInverse}), the smallest eigenvalue of $\boldsymbol M_{t+1}^{-1}$ equals to $(1-\xi^{-2})\lambda_d(\boldsymbol M_{t}^{-1} + \boldsymbol{x}_{t}\boldsymbol{x}_{t}^\top)$, which is suppressed to be small but positive. Hence, $\boldsymbol M_{t+1}^{-1}$ is guaranteed to be nonsingular. Iteratively, $\boldsymbol M_{t}^{-1}$ is positive definite for all $t\in \mathcal T$.

In the following, a theorem is given to provide a bound on the caching regret of the HPDT algorithm. Let $c$ be the caching size of each EN, and $F$ be the cardinality of the ground file set. Suppose content hit rate satisfies the linear model given by Eq. (\ref{measure}), and note that the attribute vectors are bounded as $||\boldsymbol{x}_{f,n,t}|| \leq \eta$ for all $f \in \mathcal F$, $n \in \mathcal N$ and $t\in\mathcal T$. Let $\Theta$ and $w$ be the upper bound of $||\boldsymbol{\theta}_{n,0}-\tilde{\boldsymbol{\theta}}_{n,0}||_{\boldsymbol{P}_{n,0}}^2$ and $w_{n,t}$ for all $n \in \mathcal N$ and $t \in \mathcal T$, respectively. Then, we have the following theorem.

\begin{Theorem}\label{theorem2}
The time-averaged regret $R(T)$ of the HPDT algorithm is of order $O\Big(c\eta\psi N\sqrt{\frac{\Theta}{T} + w^{2}}\Big)$.
\end{Theorem}

Please refer to Appendix \ref{prooftheorem2} for the proof. Theorem \ref{theorem2} indicates that, if the linear model is free of noises, the time-averaged regret of the HPDT algorithm tends to zero as $T$ grows to infinity. Otherwise, the HPDT algorithm may not approach the optimal solution, and its performance depends on the noise magnitude. This is due to the characteristics inherited from the $H_{\infty}$ filter. Since $H_{\infty}$ filter makes no assumption on the noise feature, to minimize the worst-case estimation error, it needs to accommodate all possible noise processes, which turns out to be over-conservative. However, when the noise is zero-mean, the regret of exploiting HPDT algorithm reduces to the order of $O(c\eta\psi N \sqrt{\frac{\Theta}{T}})$, which is smaller than that using the RPUC algorithm. In the next section, we will further evaluate the proposed two algorithms by numerically decomposing the estimation errors, and examine the algorithms by experiments on real dataset.

\section{Numerical Analysis and Experimental Results}\label{simulation}
To validate performance of the proposed caching algorithms, numerical analysis is firstly performed in this section. Afterwards, an experiment based on real-world dataset from YouTube is conducted to further illuminate the performance of the proposed algorithms in practical scenarios.

\subsection{Numerical Analysis}
The proposed two algorithms can be used for content caching with different user demand features. In essence, they are estimating the location feature vector $\boldsymbol{\theta}_{n}^{\ast}$, which specifies the location characteristics and user preferences on each EN. Given the linear model of user demand, the performance of the proposed algorithms highly depends on the accuracy of the estimation. We use mean square error (MSE) to evaluate the accuracy of the proposed algorithms. For notational simplicity, the indices $n$ and $t$ are omitted in this section. Let $\boldsymbol{\theta}$ be the underlying feature vector, and $\tilde{\boldsymbol{\theta}}$ be the estimation of $\boldsymbol{\theta}$ by using the proposed algorithms. Denote $\mathbb E [\tilde{\boldsymbol{\theta}}] = \bar{\boldsymbol{\theta}}$, then, the MSE can be defined in Euclidean norm as
\begin{eqnarray}
\mathsf{MSE}_{\tilde{\boldsymbol{\theta}}} &\!\!\!=\!\!\!& \mathbb E \big[||\tilde{\boldsymbol{\theta}} - \boldsymbol{\theta}||^2\big] \,=\, \mathbb E \Big[\big|\big|\tilde{\boldsymbol{\theta}} - \bar{\boldsymbol{\theta}} + \bar{\boldsymbol{\theta}} - \boldsymbol{\theta}\big|\big|^2\Big] \nonumber \\
%& \!\!\!=\!\!\! & \mathbb E \Big\{ \big[(\tilde{\boldsymbol{\theta}} - \bar{\boldsymbol{\theta}})^\top + (\bar{\boldsymbol{\theta}} - \boldsymbol{\theta})^\top\big] \big[(\tilde{\boldsymbol{\theta}} - \bar{\boldsymbol{\theta}}) + (\bar{\boldsymbol{\theta}} - \boldsymbol{\theta})\big]\Big\} \nonumber \\
& \!\!\!=\!\!\! &  \mathbb E \big[||\tilde{\boldsymbol{\theta}} - \bar{\boldsymbol{\theta}}||^2\big] + ||\bar{\boldsymbol{\theta}} - \boldsymbol{\theta}||^2\,.  \label{thetazero}
\end{eqnarray}
%where the last equation is due to the fact that
%\begin{eqnarray}
%\mathbb E [(\tilde{\boldsymbol{\theta}} - \bar{\boldsymbol{\theta}})^\top(\bar{\boldsymbol{\theta}} - \boldsymbol{\theta})] = \mathbb E [(\bar{\boldsymbol{\theta}} - \boldsymbol{\theta})^\top (\tilde{\boldsymbol{\theta}} - \bar{\boldsymbol{\theta}})]=\boldsymbol{0}\,.
%\end{eqnarray}
The two terms in Eq. (\ref{thetazero}) turn out to be the variance and bias of $\tilde{\boldsymbol{\theta}}$, respectively. 

Note that the RPUC algorithm is based on ridge regression. Unlike the ordinary least square linear regression, which makes an unbiased estimation on the feature vector, ridge regression intentionally introduces bias so as to reduce variance of the estimation. Moreover, the RPUC algorithm adds a perturbation to the estimation of ridge regression to account for noise uncertainty, which further increases the bias.

In contrast, the HPDT algorithm makes no assumption on the statistical model of the underlying noise process. It is able to meet the prescribed performance threshold even if the noise process leads to the worst case. Moreover, the HPDT algorithm makes unbiased estimation on the feature vector. This can be observed from the definition of the cost function $\mathcal J_0$ in Eq. (\ref{costfunction}). By decomposing the numerator of $\mathcal J_0$, we have
\begin{equation}\label{errordecompose}
\sum_{t=0}^{T}||\tilde{\boldsymbol{\theta}}_{t} - \boldsymbol{\theta}^\ast||^2 = (T+1) \cdot \mathsf{MSE}_{\tilde{\boldsymbol{\theta}}} \geq T ||\bar{\boldsymbol{\theta}} - \boldsymbol{\theta}^\ast||^2.
\end{equation}
As $H_\infty$ filter makes robust estimation over all kinds of noise structures and initial conditions, given any $\epsilon > 0$, there exists a combination of initial condition $\boldsymbol{\theta}_{0}$ and noise $\{w_{t}\}_{t=0}^T$ such that $\epsilon > ||\boldsymbol{\theta}_{0}-\tilde{\boldsymbol{\theta}}_{0} ||_{\boldsymbol{P}_0}^2 + \sum_{t=0}^{T}|w_{t}|^2 \neq 0$. Since this is true for all $\epsilon > 0$, according to Eq. (\ref{errordecompose}), the cost function $\mathcal J_0 \geq T||\bar{\boldsymbol{\theta}} - \boldsymbol{\theta}^\ast||^2/\epsilon$, which will grow linearly if $||\bar{\boldsymbol{\theta}} - \boldsymbol{\theta}^\ast||^2 \neq 0$. Consequently, any algorithm that bounds the cost function $\mathcal J_0$ must be unbiased, i.e., $||\bar{\boldsymbol{\theta}} - \boldsymbol{\theta}^\ast||^2 = 0$.

On the other hand, the performance of the HPDT algorithm also depends on the \emph{a priori} confidence on the estimation of the initial state, i.e., the selection of $\boldsymbol{P}_0$. A smaller matrix (eigenvalue) should be chosen if the estimation of initial condition is made with larger uncertainty, and vice versa.

\begin{figure}
\centering
\subfigure[]{
\includegraphics[width=0.5\textwidth]{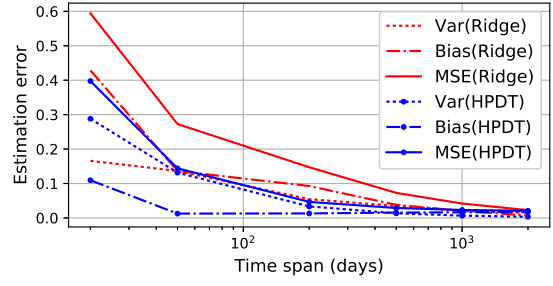}}
\subfigure[]{
\includegraphics[width=0.5\textwidth]{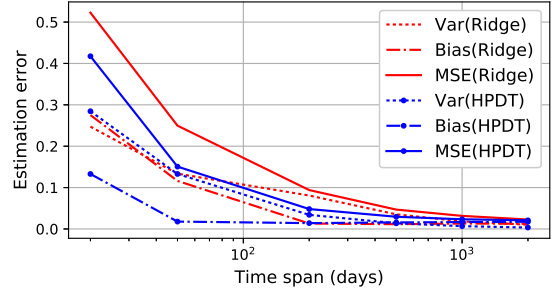}}
\subfigure[]{
\includegraphics[width=0.5\textwidth]{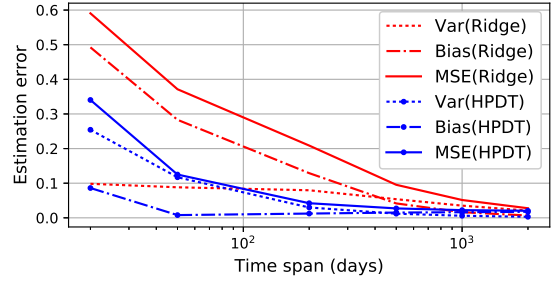}}
\caption{Comparison of estimation errors of ridge regression and HPDT algorithm under varying sample rate and noise structure. a) Zero-mean noise; b) Non-zero-mean noise of uniform distribution (with smaller mean value); c) Non-zero-mean noise of normal distribution (with larger mean value).}
\label{MSE}
\end{figure}

%\begin{figure}[t]
%\centering
%\includegraphics[width=0.45\textwidth]{}
%\caption{Comparison of estimation errors of ridge regression and HPDT algorithm under varying sample rate.}
%\label{MSE}
%\end{figure}

\begin{table*}[t]\centering
\caption{Comparison of Estimation Variance and Bias}
\bgroup
\def\arraystretch{1.3}
\begin{tabular}{c||c|c|c|c|c|c|c|c|c|c|c|c}
\hline
\multirow{2}{*}{Time Span} & \multicolumn{3}{c|}{Ridge Regression} & \multicolumn{3}{c|}{$\textrm{H}_\infty$($\lambda_{\min}(\boldsymbol{P}_0)$ = 5)} & \multicolumn{3}{c|}{$\textrm{H}_\infty$($\lambda_{\min}(\boldsymbol{P}_0)$ = 10)} & \multicolumn{3}{c}{$\textrm{H}_\infty$($\lambda_{\min}(\boldsymbol{P}_0)$ = 30)} \\
\cline{2-13}
& $\mathsf{Variance}$ & $\mathsf{Bias}$ & $\mathsf{MSE}$ & $\mathsf{Variance}$ & $\mathsf{Bias}$ & $\mathsf{MSE}$ & $\mathsf{Variance}$ & $\mathsf{Bias}$ & $\mathsf{MSE}$ & $\mathsf{Variance}$ & $\mathsf{Bias}$ & $\mathsf{MSE}$\\ \hline \hline
20 & 0.1659 & 0.4284 & 0.5943 & 0.2881 & 0.1094 & 0.3975 & 0.2687 & 0.1045 & 0.3732 & 0.2647 & 0.1036 & 0.3683\\ \hline
50 & 0.1364 & 0.1364 & 0.2728 & 0.1315 & 0.0126 & 0.1441 & 0.1244 & 0.0116 & 0.1360 & 0.1228 & 0.0115 & 0.1343\\ \hline
200 & 0.0545 & 0.0927 & 0.1472 & 0.0335 & 0.0131 & 0.0466 & 0.0318 & 0.0129 & 0.0447 & 0.0314 & 0.0129 & 0.0444\\ \hline
500 & 0.0347 & 0.0381 & 0.0728 & 0.0134 & 0.0155 & 0.0289 & 0.0127 & 0.0154 & 0.0282 & 0.0126 & 0.0154 & 0.0280 \\ \hline
1000 & 0.0211 & 0.0207 & 0.0418 & 0.0067 & 0.0164 & 0.0231 & 0.0064 & 0.0164 & 0.0227 & 0.0063 & 0.0164 & 0.0227\\ \hline
2000 & 0.0157 & 0.0070 & 0.0227 & 0.0034 & 0.0174 & 0.0207 & 0.0032 & 0.0174 & 0.0206 & 0.0032 & 0.0174 & 0.0205 \\ \hline
\end{tabular}
\egroup
\label{errortable}
\end{table*}

We conduct simulation based on synthesized time sequences, which is generated according to a prescribed linear model with zero-mean noise and non-zero-mean noise, respectively. Since the proposed RPUC algorithm is perturbed intentionally, we only present the comparison between ridge regression and the HPDT algorithm. Fig. \ref{MSE} shows that, under all scenarios, ridge regression provides a more stable estimation as it achieves smaller variance than that using HPDT algorithm when the amount of historical data is small. Such stability advantage of ridge regression benefits from its intentional penalty. However, the HPDT algorithm performs much better in terms of bias and MSE under varying sampling rate in all scenarios, which proves the robustness of the HPDT algorithm. Moreover, with the increase of noise magnitude, the gain of HPDT over ridge regression also increases.

To demonstrate the impact of $\boldsymbol{P}_0$ on the performance of HPDT, estimation results of the HPDT algorithm with different initialization matrices are presented. $\boldsymbol{P}_0$ is initialized as diagonal matrix with positive elements on the main diagonal. When the confidence on the initial state is small, a larger $\boldsymbol{P}_0$ with bigger eigenvalues is used. As shown in Table \ref{errortable}, $\boldsymbol{P}_0$ with larger eigenvalue achieves better performance than the others, which means that the initial guess $\boldsymbol{\theta}_0$ is close to the prescribed vector. In practice, the matrix $\boldsymbol{P}_0$ can be selected according to the prior information regarding the initial condition.

\subsection{Real Dataset Experiment}
\begin{figure}
\centering
\includegraphics[width=0.6\textwidth]{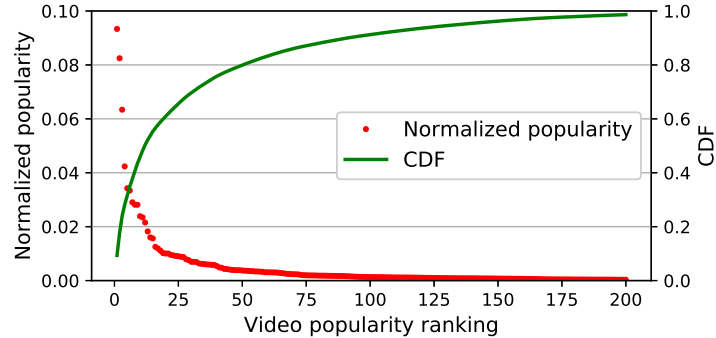}
\caption{Popularity skewness of the video set in our experiment. Note that the videos are randomly crawled from YouTube, which may not reflect the overall skewness of video popularity.}
\label{skewness}
\end{figure}

\begin{figure*}
\centering
\subfigure[]{\label{c=5}
\includegraphics[width=0.5\textwidth]{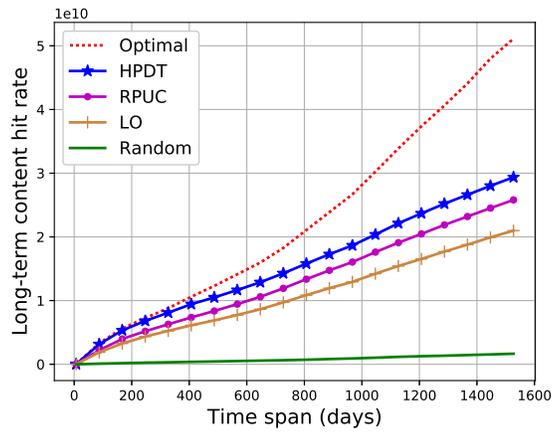}}
\subfigure[]{\label{c=20}
\includegraphics[width=0.5\textwidth]{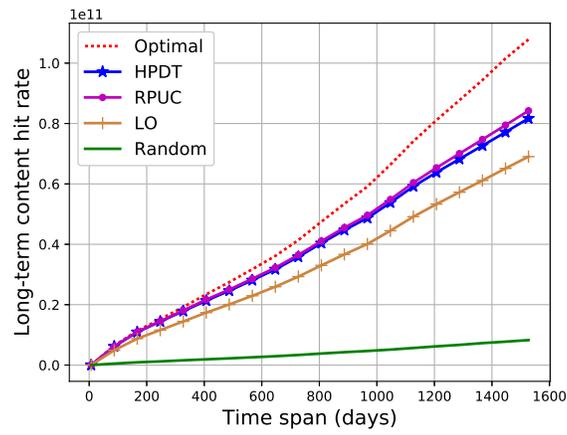}}
\subfigure[]{\label{c=100}
\includegraphics[width=0.5\textwidth]{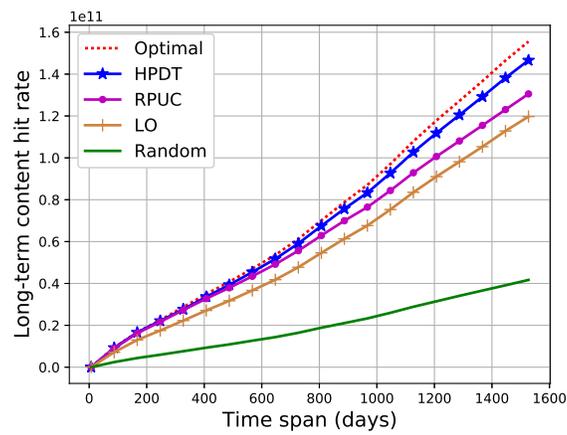}}
\caption{The content hit rate comparison between the proposed algorithm and other benchmarks with varying caching size, where the total number of files is 800 and the caching sizes of six figures are 5, 20, and 100 respectively.}
\label{simresult}
\end{figure*}

\subsubsection{Experiment Setup}
To further demonstrate the advantages of the proposed algorithms in practical scenarios, we conduct an experiment on the dataset crawled from YouTube. On YouTube, some video owners made their video view statistics open to public. Among other information, the view amount information is recorded on a daily basis. To obtain such information, a Python-based crawling program is written, and the request record of each video is crawled into a \texttt{.json} file. Based on which we conducted the rest Python-based experiments . In total, $800$ videos are randomly crawled, which were uploaded before January 2013, with full view statistics till May 2017. The most popular video has been watched over $2.85$ billion times by the end of the timespan, while the least popular one has been rarely viewed across the time span. Fig. \ref{skewness} shows the statistics of video popularity skewness. The popularity of the most popular $25$ videos is highly skewed, and the most popular 50 videos account for almost $80\%$ of the total view amount.

Note that the dataset only contains the global statistics of each video record (with the recent update of YouTube webpage, even the global view statistics is inaccessible), while the view statistics of most online video content providers in a local area is unavailable. To emulate the video request processes on different locations, the original statistic of each video is shifted and scaled randomly over the time span. For the record of a certain video, by shifting, the request record of the original global data of each content is moved backward and forward on the timespan. By scaling, each content request record is then randomly scaled up and down. After shifting and scaling, the original global request statistics are transformed and treated as the requests from different locations. The key point is that the pattern of the record remains valid after the above transformation. In this way, we are able to characterize the location diversity based on the emulated view statistics.

Specifically, consider the content library containing those $800$ videos, each video can be cached on $3$ ENs, each with caching size $c$. Content refreshing is performed upon the network traffic pattern. For example, wireless traffic presents regular peak and valley every day. Hence, content refreshing can be performed during the off-peak period with minimal impact on normal network activity. A video can be characterized from several aspects, including video quality, genre, length, and historical view statistic. In this experiment, we use view amount information in the past $7$ days as the attribute vector of each content, i.e., $d = 7$. Based on the attribute vectors and an initial guess on the content popularity, the algorithms gradually select contents that are predicted to be more popular than the others, and cache them on each EN accordingly. The long-term content hit rates of the proposed algorithms are shown by comparing with the following benchmark algorithms. 1) Hindsight optimal. By analyzing the full view record over the time span at each EN, the most popular videos are selected and cached respectively. Note that this benchmark requires future information and hence cannot be implemented in practice. 2) Location oblivious (denoted by LO). During each time slot, the historical demands of all the contents from all ENs are analyzed, afterwards the ones that are predicted (by ridge regression) to have the highest demands in the next time slot are identified. Then, all ENs will cache the same set of contents without location differentiation. 3) Random. A random set of videos is selected to update the ENs during each time slot.

\subsubsection{Experimental Results}
As shown in Fig. \ref{skewness}, the popularity of YouTube video is highly skewed, and the most popular $10\%$ videos have attracted almost $90\%$ of user requests. The skewness of video popularity has also been validated in \cite{imc}. The popularity of this dataset can be roughly divided into three levels: highly skewed (popularity of the top $15$ videos); medium skewed (popularity of videos ranking from $16$ to $50$); and less skewed (the rest ones).

Figure \ref{simresult} shows the comparison of long-term content hit rates of different algorithms with varying EN caching sizes. The performance of the caching algorithms is affected by the skewness of the popularity profile. However, the proposed location-based approaches always outperform the location-oblivious scheme in varying caching size scenarios. Specifically, when the caching size falls into the highly skewed area and less skewed area, the proposed caching algorithms RPUC and HPDT outperform other benchmarks considerably. In particular, the HPDT algorithm performs better than the RPUC. For the highly skewed area, the top $5$ videos present much higher variance than the rest. As a result, the noise mean of those records is also significant. Since RPUC algorithm is designed for zero-mean noises, their performance is limited when noise amplitude is significant. In contrast, when the caching size falls into the less skewed area, the algorithms need to corporate various noise types of different videos, which may not always be zero-mean. RPUC and HPDT perform equally well when the caching size falls into the medium skewed area (Fig. \ref{c=20}). The $H_\infty$ filter is utilized to provide guaranteed performance even when noise type lead to the worst case for estimation. As a result, the HPDT algorithm is conservative yet robust.

Figure \ref{skewness} also indicates that content popularity is long-tailed, i.e., the less popular contents attract almost vanishing requests compared to the popular ones. As a result, the total hit rate of different caching schemes in Fig. \ref{simresult} does not increase linearly with the caching size. Note that, both algorithms run iteratively in an online fashion. During each iteration, the most computationally intensive execution is the $n$ times sorting of the estimated demands of $F$ contents, which has a typical computational complexity of $O(nF log F)$. Complexity of the value assignments and matrix update could be neglected compared with sorting. As a result, both algorithms are of low time complexity.

\subsection{Discussions}
\subsubsection{Another dimension of the prediction}
This work focuses on the estimation of location feature vector. Actually, the selection of video attribute vector $\boldsymbol{x}_{f,n}$ also influences the prediction accuracy. As mentioned before, other factors, such as video quality, length and genre, can also be used to characterize video contents. If such labeling information is available, by reducing the dimension as well as training the dataset, we can identify influential features that affect content popularity.
On the other hand, for the location feature vector $\boldsymbol{\theta}_{n}^{\ast}$, both RPUC and HPDT algorithms are designed with the precondition that $\boldsymbol{\theta}_{n}^{\ast}$ is time-invariant, as indicated in both Eq. (\ref{linearprediction}) and (\ref{measure}). Actually, the HPDT algorithm can be directly extended to time-variant scenario if the state equation is also linear, i.e., $\boldsymbol{\theta}_{t+1} = \boldsymbol{A} \boldsymbol{\theta}_{t} + v_{t}$, where $\boldsymbol{A} \in \mathbb R^{d\times d}$ is the transition matrix, and $v_{t} \in \mathbb R^{d}$ is the state noise vector. By resorting to $H_{\infty}$ technique, the adaptive estimation on $\boldsymbol{\theta}_{n,t}$ can be made with guaranteed accuracy \cite{shen}.

\subsubsection{The applicability in practical video streaming}
The proposed popularity prediction approach can be applied to the delivery of various types of contents. As video content consumes the most bandwidth, it deserves to be in-depth investigated. Practical video streaming protocols (such as HTTP-based Adaptive Streaming, HAS) divide a video content into several chunks/segments, each with multiple bitrates and quality versions \cite{tmm18}. Those pull-based streaming protocols dynamically change the quality of the streamed video according to the observed network conditions on a per-fragment basis. Most of the research works on adaptive video streaming (both server-side bitrate switching \cite{nossdav13} and client-side switching \cite{sigcomm15}) strive to predict the network condition when transmitting the next video segment. In contrast, with the aid of edge storage resources, our work focuses on push-based content distribution. In other words, estimating the available bandwidth is out of the scope of this work, and content updates are scheduled to off-peak periods, where streaming bandwidth is sufficient. 

When streaming video contents based on HAS, whether to cache individual segments or the whole quality representation depends on both the available bandwidth and the content popularity. In particular, for the popular contents, users tend to keep requesting them regardless of the received video quality. Hence, it is more appropriate to store the whole representation of the video so as to fulfill users' requirements via dynamic bandwidth. As the popularity profile of contents are highly skewed (shown in Fig. \ref{skewness}), the streaming provider only needs to store the full quality representation of the most popular videos (the amount of which really depends on the caching size budget). For the less popular ones, the provider may choose to cache the individual segments that are of moderate quality, so as to save bandwidth and meanwhile be responsive to user requests. The rationale behind such decision is the content popular profile and the available caching resources, which is the merit of our work. In this sense, the proposed location-based popularity prediction approaches are crucial in HAS-based streaming system, and the prediction of content popularity and network condition will collectively contribute to improved video streaming.

\section{Conclusion}\label{conclusion}
In this paper, we investigate popularity prediction for mobile edge caching, with special focus on location awareness. We model the content popularity profile by a linear model and propose online algorithms to deal with different statistical models of the noise process. The proposed RPUC algorithm achieves content hit rate that asymptotically approaches the optimal solution when the noise is zero-mean. Noticing that the noise may not necessarily be zero-mean, we resort to the $H_\infty$ filter technique and propose the HPDT algorithm for popularity prediction. This algorithm can achieve guaranteed prediction accuracy even when the worst-case noise occurs. Both algorithms can be implemented without training phases. Numerical analysis shows how the performance of the proposed algorithms is affected by different types of noises, the amount of historical data, and the initial state. Extensive experiments on real dataset demonstrate the advantage of the proposed algorithm, which helps to make customized caching decisions in practical scenarios. For future works, we will exploit locational features of neighboring ENs to make better caching decisions.

\appendices
\section{Proof of Lemma \ref{estimationerror}} \label{prooflemma1}
Let $\boldsymbol{h}_{f,n} = \boldsymbol{\Phi}_{f,n}^\top \boldsymbol{y}_{f,n}$, based on Eq. (\ref{ridge}), the estimation error can be rewritten as
\begin{eqnarray}
&&|\boldsymbol{x}_{f,n}^\top \tilde{\boldsymbol{\theta}}_n - \boldsymbol{x}_{f,n}^\top \boldsymbol{\theta}_n^\ast| \nonumber\\
&=& |\boldsymbol{x}_{f,n}^\top \boldsymbol{V}_{f,n}^{-1} \boldsymbol{h}_{f,n} - \boldsymbol{x}_{f,n}^\top \boldsymbol{V}_{f,n}^{-1} (\boldsymbol{\Phi}_{f,n}^\top \boldsymbol{\Phi}_{f,n} + \mu \boldsymbol{I}_d) \boldsymbol{\theta}_n^\ast| \nonumber \\
&=& |\boldsymbol{x}_{f,n}^\top \boldsymbol{V}_{f,n}^{-1} \boldsymbol{\Phi}_{f,n}^\top (\boldsymbol{y}_{f,n} - \boldsymbol{\Phi}_{f,n}\boldsymbol{\theta}_n^\ast) - \mu \boldsymbol{x}_{f,n}^\top \boldsymbol{V}_{f,n}^{-1} \boldsymbol{\theta}_n^\ast|. \nonumber
\end{eqnarray}
Since $||\boldsymbol{\theta}_n^\ast|| \leq \zeta$, H\"older's inequality indicates that $|\mu \boldsymbol{x}_{f,n}^\top \boldsymbol{V}_{f,n}^{-1} \boldsymbol{\theta}_n^\ast| \leq \zeta\mu ||\boldsymbol{x}_{f,n}^\top \boldsymbol{V}_{f,n}^{-1}||$. Then, the estimation error is bounded as
\begin{eqnarray}
|\boldsymbol{x}_{f,n}^\top \tilde{\boldsymbol{\theta}}_n - \boldsymbol{x}_{f,n}^\top \boldsymbol{\theta}_n^\ast| & \!\! \leq \!\! & |\boldsymbol{x}_{f,n}^\top \boldsymbol{V}_{f,n}^{-1} \boldsymbol{\Phi}_{f,n}^\top (\boldsymbol{y}_{f,n} - \boldsymbol{\Phi}_{f,n}\boldsymbol{\theta}_n^\ast)|\nonumber\\
&& + \;\; \zeta\mu ||\boldsymbol{x}_{f,n}^\top \boldsymbol{V}_{f,n}^{-1}||. \label{twoerror}
\end{eqnarray}
The right-hand side of above inequality decomposes the estimation error into two parts, where the first (variance term) specifies the error caused by linear model, and the second (bias term) is the bias incurred by ridge regression parameter $\mu$. According to Eq. (\ref{linearprediction}), we have $\mathbb E [\boldsymbol{y}_{f,n} - \boldsymbol{\Phi}_{f,n}\boldsymbol{\theta}_n^\ast] = 0$. The Azuma's inequality gives a probabilistic upper bound on the variance term of Eq. (\ref{twoerror}):
\begin{eqnarray}
&&\mathbb{P}\bigg\{\Big|\boldsymbol{x}_{f,n}^\top \boldsymbol{V}_{f,n}^{-1} \boldsymbol{\Phi}_{f,n}^\top (\boldsymbol{y}_{f,n} - \boldsymbol{\Phi}_{f,n}\boldsymbol{\theta}_n^\ast)\Big| > \delta ||\boldsymbol{x}_{f,n}||_{\boldsymbol{V}_{f,n}^{-1}}\bigg\} \nonumber \\
&&\leq \,\, 2\exp\Big(-\frac{2\delta^2 ||\boldsymbol{x}_{f,n}||_{\boldsymbol{V}_{f,n}^{-1}}^2} {||\boldsymbol{x}_{f,n}^\top \boldsymbol{V}_{f,n}^{-1} \boldsymbol{\Phi}_{f,n}^\top||^2}\Big) \,\,\leq \,\,2e^{-2\delta^2}, \label{error1}
\end{eqnarray}
where the last inequality is due to the fact that
\begin{eqnarray}
||\boldsymbol{x}_{f,n}||_{\boldsymbol{V}_{f,n}^{-1}}^2 &=& \boldsymbol{x}_{f,n}^\top \boldsymbol{V}_{f,n}^{-1} (\boldsymbol{\Phi}_{f,n}^\top \boldsymbol{\Phi}_{f,n} + \mu \boldsymbol{I}_d) \boldsymbol{V}_{f,n}^{-1} \boldsymbol{x}_{f,n} \nonumber \\
& \geq & \boldsymbol{x}_{f,n}^\top \boldsymbol{V}_{f,n}^{-1} \boldsymbol{\Phi}_{f,n}^\top \boldsymbol{\Phi}_{f,n} \boldsymbol{V}_{f,n}^{-1} \boldsymbol{x}_{f,n} \nonumber \\
& = & ||\boldsymbol{x}_{f,n}^\top \boldsymbol{V}_{f,n}^{-1} \boldsymbol{\Phi}_{f,n}^\top||^2.
\end{eqnarray}
Hence, the variance term of Eq. (\ref{twoerror}) can be bounded by $\delta ||\boldsymbol{x}_{f,n}||_{\boldsymbol{V}_{f,n}^{-1}}$ with probability at least $1-2e^{-2\delta^2}$. Further, The bias term of Eq. (\ref{twoerror}) can be bounded as
\begin{eqnarray}
||\boldsymbol{x}_{f,n}^\top \boldsymbol{V}_{f,n}^{-1}|| & = & \sqrt{\boldsymbol{x}_{f,n}^\top \boldsymbol{V}_{f,n}^{-1} \boldsymbol{I}_d \boldsymbol{V}_{f,n}^{-1} \boldsymbol{x}_{f,n}} \nonumber \\
& \leq & \sqrt{\boldsymbol{x}_{f,n}^\top \boldsymbol{V}_{f,n}^{-1} (\mu\boldsymbol{I}_d+ \boldsymbol{\Phi}_{f,n}^\top \boldsymbol{\Phi}_{f,n}) \boldsymbol{V}_{f,n}^{-1} \boldsymbol{x}_{f,n}} \nonumber \\
& = & ||\boldsymbol{x}_{f,n}||_{\boldsymbol{V}_{f,n}^{-1}}. \label{error2}
\end{eqnarray}
By substituting Eq. (\ref{error1}) and (\ref{error2}) into Eq. (\ref{twoerror}), the probabilistic bound in Eq. (\ref{errorbound}) directly follows.

\section{Proof of Theorem 1}\label{prooftheorem1}
The total regret depends on the RUPC algorithm's accuracy of estimation on content hit rate, which is elaborated in Lemma \ref{estimationerror}. According to this lemma, the true hit rate of file $f$ at EN $n$ lies in the confidence interval around the predicted hit rate
\begin{equation}
\mathcal I_{f,n,t} = [\boldsymbol{x}_{f,n,t}^\top \tilde{\boldsymbol{\theta}}_{n,t} - p_{f,n,t}, \; \boldsymbol{x}_{f,n,t}^\top \tilde{\boldsymbol{\theta}}_{n,t} + p_{f,n,t}]
\end{equation}
with high probability.

Let $\mathcal X_{n,t} = \{\exists f \in\mathcal F: |d_{f,n,t} - \tilde d_{f,n,t}| \geq p_{f,n,t}\}$ be the event that there exists at least one file whose true hit rate lies outside its confidence interval. Let $\bar{\mathcal X}_{n,t}$ be the complementary event of $\mathcal X_{n,t}$, i.e., all files' true hit rates fall into their confidence interval. Let $r_{n,t}$ be the instant regret of a caching algorithm in EN $n$ at time slot $t$. According to Eq. (\ref{regret}), the total regret depends on the difference between the set of files chosen by the Algorithm and the optimum set, i.e., $\mathcal F_{n,t}$ and $\mathcal F_{n,t}^\ast$, thus
\begin{equation}\label{instantregret}
\begin{array}{c}
r_{n,t} = \sum_{f \in \mathcal F_{n,t}^\ast} d_{f,n,t} - \sum_{f \in \mathcal F_{n,t}} d_{f,n,t},
\end{array}
\end{equation}
and the time-averaged regret can be rewritten as
\begin{eqnarray}
R(T) &\!\!\!\!\!=\!\!\!\!\!& \frac{1}{T}  \sum_{t\in\mathcal T} \sum_{n \in \mathcal N} r_{n,t} \nonumber\\
&\!\!\!\!\!=\!\!\!\!\!& \frac{1}{T} \sum_{t\in\mathcal T} \sum_{n \in \mathcal N} \mathbbm 1_{\{\mathcal X_{n,t}\}}r_{n,t} + \frac{1}{T}  \sum_{t\in\mathcal T} \sum_{n \in \mathcal N} \mathbbm 1_{\{\bar{\mathcal X}_{n,t}\}} r_{n,t}, \ \ \ \ \ \label{dividedregret}
\end{eqnarray}
where $\mathbbm 1_{\{\mathcal X_{n,t}\}}$ is an indicator variable that equals to $1$ if event $\mathcal X_{n,t}$ happens, and equals to $0$ otherwise. To proceed, the two terms in Eq. (\ref{dividedregret}) are bounded separately.

Firstly, consider the case when event $\mathcal X_{n,t}$ happens. With the setting of $\alpha_t$ in Eq. (\ref{perturbation}), for a file $f$ and EN $n$ at time $t$, we have $\mathbb P\{|d_{f,n,t} - \tilde d_{f,n,t}|\geq p_{f,n,t}\} \leq 2F^{-1}t^{-2}$. As a result, the frequency of event $\mathcal X_{n,t}$ happens on all ENs over the time span can be bounded as:
\begin{eqnarray}
\sum_{t\in\mathcal T} \sum_{n \in \mathcal N} \mathbbm 1_{\{\mathcal X_{n,t}\}} &\!\!\!\leq\!\!\!& \sum_{t\in\mathcal T} \sum_{n \in \mathcal N} \sum_{f \in \mathcal F} \mathbb P\big\{|d_{f,n,t} - \tilde d_{f,n,t}|\geq p_{f,n,t}\big\} \nonumber\\
&\!\!\!\leq\!\!\!& \sum_{t\in\mathcal T} \sum_{n \in \mathcal N} \sum_{f \in \mathcal F} 2F^{-1}t^{-2} = 2N \sum_{t\in\mathcal T} t^{-2} \nonumber\\
&\!\!\!\leq\!\!\!& 2N \sum_{t=1}^\infty t^{-2} \leq \frac{\pi^2}{3}N.
\end{eqnarray}
As content hit rate $d_{f,n,t}\leq \gamma$, according to Eq. (\ref{instantregret}), a coarse upper bound of the instant regret is $r_{n,t} \leq c \gamma$, where $c = |\mathcal F_{n,t}^\ast|$ is the caching size of each EN. Therefore, the first term of Eq. (\ref{dividedregret}) can be bound as
\begin{equation}\label{firstbound}
\frac{1}{T}  \sum_{t\in\mathcal T} \sum_{n\in\mathcal N} \mathbbm 1_{\{\mathcal X_{n,t}\}} r_{n,t} \leq \frac{\pi^2 c\gamma N}{3T}. 
\end{equation}

Then, consider the case when event $\bar{\mathcal X}_{n,t}$ happens, all files' true hit rates falls into the confidence interval around their estimation $\tilde d_{f,n,t}$. Hence, $|d_{f,n,t} - \tilde d_{f,n,t}|\leq p_{f,n,t}, \; \forall f\in\mathcal F$. With $\hat d_{f,n,t} = \tilde d_{f,n,t} + p_{f,n,t}$, we have
\begin{equation}\label{confidence}
\hat d_{f,n,t} - d_{f,n,t} \leq 2p_{f,n,t}.
\end{equation}
By Eq. (\ref{instantregret}) and (\ref{confidence}), when event $\bar{\mathcal X}_{n,t}$ happens, the instant regret $r_{n,t}$ can be bounded as
\begin{equation}
\begin{array}{rcl}
r_{n,t}|_{\bar{\mathcal X}_{n,t}} &=& \sum_{f \in \mathcal F_{n,t}^\ast \setminus \mathcal F_{n,t}} d_{f,n,t} - \sum_{f \in \mathcal F_{n,t} \setminus \mathcal F_{n,t}^\ast} d_{f,n,t} \\
&\leq& \sum_{f \in \mathcal F_{n,t}^\ast \setminus \mathcal F_{n,t}} \hat d_{f,n,t} - \sum_{f \in \mathcal F_{n,t} \setminus \mathcal F_{n,t}^\ast} d_{f,n,t} \\
&\leq& \sum_{f \in \mathcal F_{n,t} \setminus \mathcal F_{n,t}^\ast} \Big(\hat d_{f,n,t} - d_{f,n,t}\Big) \label{setfunction}\\
&\leq& 2\sum_{f \in \mathcal F_{n,t} \setminus \mathcal F_{n,t}^\ast} p_{f,n,t},
\end{array}
\end{equation}
where inequality (\ref{setfunction}) is due to fact that since the algorithm selects files in $\mathcal F_{n,t} \setminus \mathcal F_{n,t}^\ast$ rather than $\mathcal F_{n,t}^\ast \setminus \mathcal F_{n,t}$, hence the collective upper confidence bound hit rate satisfies $\sum_{f \in \mathcal F_{n,t} \setminus \mathcal F_{n,t}^\ast} \hat d_{f,n,t} \geq \sum_{f \in \mathcal F_{n,t}^\ast \setminus \mathcal F_{n,t}} \hat d_{f,n,t}$.
We will need two lemmas from \cite{Yasin} for the subsequent analysis.

\begin{Lemma}\label{lemma11}
(Lemma 11 of \cite{Yasin}). Let $\boldsymbol{x}_1,\boldsymbol{x}_2,\cdots,\boldsymbol{x}_T \in\mathbb R^d$ and $\boldsymbol{V}_t = \mu\boldsymbol{I}_d + \sum_{s=1}^t \boldsymbol{x}_s\boldsymbol{x}_s^\top$. If $\forall t\in\mathcal T$, $||\boldsymbol{x}_t||\leq \eta$ holds, then for some $\mu>0$, when $\mu \geq \max(1,\eta^2)$, we have
\begin{equation}
\sum_{s=1}^t (\boldsymbol{x}_s^\top \boldsymbol{V}_t^{-1} \boldsymbol{x}_s) \leq 2\ln\frac{\det(\boldsymbol{V}_t)}{\mu}.
\end{equation}
\end{Lemma}

\begin{Lemma}\label{lemma10}
(Determinant-Trace Inequality, Lemma 10 of \cite{Yasin}). Suppose $\boldsymbol{x}_1,\boldsymbol{x}_2,\cdots,\boldsymbol{x}_T \in\mathbb R^d$, and $\forall t\in\mathcal T$ it holds that $||\boldsymbol{x}_t||\leq \eta$. Let $\boldsymbol{V}_t = \mu\boldsymbol{I}_d + \sum_{s=1}^t \boldsymbol{x}_s\boldsymbol{x}_s^\top$, then for $\mu>0$, we have
\begin{equation}
\det(\boldsymbol{V}_t) \leq (\mu + t \eta^2/d)^d.
\end{equation}
\end{Lemma}

Based on those two lemmas, the second term in Eq. (\ref{dividedregret}) can be bounded as
\begin{eqnarray}
&&\sum_{t\in\mathcal T} \sum_{n\in\mathcal N}  r_{n,t}|_{\bar{\mathcal X}_{n,t}} \;\;\leq\;\;2\sum_{t\in\mathcal T} \sum_{n \in\mathcal N}\sum_{f \in \mathcal F_{n,t}\setminus \mathcal F_{n,t}^\ast} p_{f,n,t}\nonumber\\
&\leq& 2c\alpha_T \sum_{n\in\mathcal N} \sum_{t\in\mathcal T} ||\boldsymbol{x}_{f,n,t}||_{\boldsymbol{V}_{f,n}^{-1}}\label{midalpha}\\
&\leq& 2c\alpha_T \sum_{n\in\mathcal N} \sqrt{T\sum_{t\in\mathcal T} ||\boldsymbol{x}_{f,n,t}||_{\boldsymbol{V}_{f,n}^{-1}}^2}\label{midroot}\\
&\leq& 2c\alpha_TN \sqrt{2T \ln\frac{(\mu + T\eta^2/d)^d}{\mu}},\label{midlemmas}
\end{eqnarray}
where Eq. (\ref{midalpha}) is due the fact that $\alpha_t$ increases with $t$, Eq. (\ref{midroot}) holds because the arithmetic mean of a set of values is smaller than their root-mean square and Eq. (\ref{midlemmas}) is based on the above two lemmas.

By substituting Eq. (\ref{midlemmas}) and (\ref{firstbound}) into Eq. (\ref{dividedregret}), and together with $\alpha_T = \sqrt{\ln (TF^{\frac{1}{2}})} + \mu\zeta$, we have
\begin{eqnarray}
R(T) &\leq& 2c\alpha_TN\sqrt{\frac{2}{T}\ln\frac{(\mu + T \eta^2/d)^d}{\mu}} + \frac{\pi^2c\gamma N}{3T} \nonumber\\
& = & O\Big(cN\sqrt{\frac{d (\ln T) \ln(\mu + T\eta^2/d)}{T}}\Big).
\end{eqnarray}

\section{Proof of Theorem 2}\label{prooftheorem2}
The typical estimation error on the demand of a certain content $f$ on EN $n$ during time slot $t$ is $|d_{f,n,t} - \tilde d_{f,n,t}|$. Based on H\"older's inequality, the estimation error can be bounded as
\begin{eqnarray}
&& |d_{f,n,t} - \tilde d_{f,n,t}| = |\boldsymbol{x}_{f,n,t}^\top \tilde{\boldsymbol{\theta}}_{n,t} - \boldsymbol{x}_{f,n,t}^\top \boldsymbol{\theta}_{n}^\ast| \nonumber\\
&\leq \!\!\!&  ||\boldsymbol{x}_{f,n,t}||\cdot||\tilde{\boldsymbol{\theta}}_{n,t}-\boldsymbol{\theta}_{n}^\ast|| \leq  \eta ||\tilde{\boldsymbol{\theta}}_{n,t}-\boldsymbol{\theta}_{n}^\ast||. \label{holder}
\end{eqnarray}
Define the per-EN per-slot regret as
\begin{equation}
\begin{array}{c}
r_{n,t} = \sum_{f \in \mathcal F_{n,t}^\ast} d_{f,n,t} - \sum_{f \in \mathcal F_{n,t}} d_{f,n,t}.
\end{array}
\end{equation}
Since the files in $\mathcal F_{n,t} \setminus \mathcal F_{n,t}^\ast$ is selected rather than the ones in $\mathcal F_{n,t}^\ast \setminus \mathcal F_{n,t}$, the estimated user demands of those files satisfy 
\begin{equation}\label{diff}
\begin{array}{c}
\sum_{f \in \mathcal F_{n,t}^\ast \setminus \mathcal F_{n,t}} \tilde d_{f,n,t} \leq \sum_{f \in \mathcal F_{n,t} \setminus \mathcal F_{n,t}^\ast} \tilde d_{f,n,t}.
\end{array}
\end{equation}
Eq. (\ref{holder}) indicates that $\tilde{d}_{f,n,t}\leq d_{f,n,t}+\eta ||\tilde{\boldsymbol{\theta}}_{n,t}-\boldsymbol{\theta}_{n}^\ast||$ and $d_{f,n,t}\leq \tilde{d}_{f,n,t}+\eta ||\tilde{\boldsymbol{\theta}}_{n,t}-\boldsymbol{\theta}_{n}^\ast||$ for all $f\in\mathcal F$, $n\in\mathcal N$ and $t\in\mathcal T$. With Eq. (\ref{diff}), we have
\begin{equation}
\begin{array}{rcl}
r_{n,t} &=& \sum_{f \in \mathcal F_{n,t}^\ast\setminus\mathcal F_{n,t}} d_{f,n,t} - \sum_{f \in \mathcal F_{n,t}\setminus\mathcal F_{n,t}^\ast} d_{f,n,t}\nonumber\\
&\leq& \sum_{f \in \mathcal F_{n,t}^\ast\setminus\mathcal F_{n,t}} (\tilde d_{f,n,t}+\eta ||\tilde{\boldsymbol{\theta}}_{n,t}-\boldsymbol{\theta}_{n}^\ast||) \nonumber\\
&& - \sum_{f \in \mathcal F_{n,t}\setminus\mathcal F_{n,t}^\ast} (\tilde d_{f,n,t}-\eta ||\tilde{\boldsymbol{\theta}}_{n,t}-\boldsymbol{\theta}_{n}^\ast||) \nonumber\\
&\leq& \sum_{f \in \mathcal F_{n,t}^\ast\setminus\mathcal F_{n,t}} \eta ||\tilde{\boldsymbol{\theta}}_{n,t}-\boldsymbol{\theta}_{n}^\ast|| \nonumber\\
&& + \sum_{f \in \mathcal F_{n,t}\setminus\mathcal F_{n,t}^\ast} \eta ||\tilde{\boldsymbol{\theta}}_{n,t}-\boldsymbol{\theta}_{n}^\ast|| \nonumber\\
&\leq& 2c\eta ||\tilde{\boldsymbol{\theta}}_{n,t}-\boldsymbol{\theta}_{n}^\ast||.
\end{array}
\end{equation}

Then, the cumulative mean regret can be rewritten as
\begin{eqnarray}
R(T) &=& \frac{1}{T} \sum_{t\in\mathcal T}\sum_{n\in\mathcal N} r_{n,t} \leq \frac{2c\eta}{T} \sum_{n\in\mathcal N} \sum_{t\in\mathcal T}||\tilde{\boldsymbol{\theta}}_{n,t}-\boldsymbol{\theta}_{n}^\ast|| \nonumber\\
&\leq& \frac{2c\eta}{T} \sum_{n\in\mathcal N} \sqrt{ T \sum_{t\in\mathcal T}||\tilde{\boldsymbol{\theta}}_{n,t}-\boldsymbol{\theta}_{n}^\ast||^{2}}. \label{regret1}
\end{eqnarray}
Since the initialization error and noises are bounded as $||\boldsymbol{\theta}_{n,0}-\tilde{\boldsymbol{\theta}}_{n,0}||_{\boldsymbol{P}_{n,0}}^2 \leq \Theta$ and $|w_{n,t}| \leq w$ for all $t \in \mathcal T$ and $n \in \mathcal N$. Based on Eq. (\ref{costfunction}) and (\ref{costthreshold}), for a certain EN $n$, we have 
\begin{eqnarray}
\sum_{t\in\mathcal T}||\tilde{\boldsymbol{\theta}}_{n,t}-\boldsymbol{\theta}_{n}^\ast||^{2} &\!\!\!\leq \!\!\!& \psi^{2} (||\boldsymbol{\theta}_{n,0}-\tilde{\boldsymbol{\theta}}_{n,0}||_{\boldsymbol{P}_{n,0}}^2 + \sum_{t \in \mathcal T} |w_{n,t}|^2) \nonumber\\
&\!\!\! \leq \!\!\!& \psi^{2} (\Theta + T w^{2}). \label{thetabound}
\end{eqnarray}
By substituting Eq. (\ref{thetabound}) into Eq. (\ref{regret1}), the time-averaged caching regret is finally bounded as:
\begin{equation}
R(T) \leq 2c\eta\psi N \sqrt{\frac{\Theta}{T} + w^{2}},
\end{equation}
which completes the proof.

%\section*{Acknowledgment}
%This work is supported by National Natural Science Foundation of China under Grant No. 61231010, Research Fund for the Doctoral Program of MOE of China under Grant No. 20120142110015 and the Natural Sciences and Engineering Research Council (NSERC) of Canada. Peng Yang is also financially supported by the China Scholarship Council.

\end{document}